\documentclass[conference]{IEEEtran}
\IEEEoverridecommandlockouts
\usepackage{cite}
\usepackage{amsmath,amssymb,amsfonts}
\usepackage{algorithmic}
\usepackage{graphicx}
\usepackage{textcomp}
\usepackage{xcolor}

\usepackage{booktabs}
\usepackage{makecell}
\usepackage{multirow}

\usepackage{etoolbox}
\makeatletter
\patchcmd{\@makecaption}
  {\scshape}
  {}
  {}
  {}
\makeatother
\usepackage[acronym]{glossaries}
\newacronym{ORNL}{ORNL}{Oak Ridge National Laboratory}
\newacronym{OLCF}{OLCF}{Oak Ridge Leadership Computing Facility}
\newacronym{Trento}{EPYC™ 7A53 “Trento”}{{\color{red}This should not show!}}
\glsunset{Trento}%To only show the Abbreviation
\newacronym{MI250}{Instinct™ MI250X}{{\color{red}This should not show!}}
\glsunset{MI250}
\newacronym{GCD}{GCD}{Graphics Compute Die}
\newacronym{NIC}{NIC}{Network Interface Card}
\newacronym{CPU}{CPU}{Central Processing Unit}
\glsunset{CPU}
\newacronym{GPU}{GPU}{Graphics Processing Unit}
\glsunset{GPU}
\newacronym{TDP}{TDP}{Thermal Design Power}
\newacronym{ALU}{ALU}{Arithmetic Logic Unit}
\newacronym{GPP}{GPP}{General Plasmon Pole}
\newacronym{FMA}{FMA}{Fused Multiply Add}
\newacronym{ERT}{ERT}{Empirical Roofline Tool}
\newacronym{DVFS}{DVFS}{Dynamic Voltage and Frequency Scaling}
\newacronym{SIMD}{SIMD}{Single Instruction, Multiple Data}
\newacronym{FLOPS/s}{FLOPS/s}{FLoating point Operations Per Second}
\newacronym{ODA}{ODA}{Operational Data Analysis}
\newacronym{HBM2e}{HBM2e}{on-die High Bandwidth Memory}
\newacronym{VAI}{VAI}{Variable Arithmetic Intensity}
\glsdisablehyper

\usepackage{siunitx}
\usepackage[
  plainruled,
  linesnumbered,
  noline
]{algorithm2e}
\makeatletter
\def\@algocf@capt@plainruled{above}
\renewcommand{\algocf@caption@plainruled}{%
  \vskip\AlCapSkip%
  \box\algocf@capbox%
  \vskip 5\algoheightrule}%
\makeatother
\usepackage{comment}
\def\BibTeX{{\rm B\kern-.05em{\sc i\kern-.025em b}\kern-.08em
    T\kern-.1667em\lower.7ex\hbox{E}\kern-.125emX}}
\begin{document}

%% For making comments in the text 
%\usepackage{color}
%\usepackage[dvipsnames]{xcolor}

\newcommand{\loremipsum}[1]{\noindent{{\color{gray}{\sc \bf LOREMIPSUM:}   #1}}}
\newcommand{\rev}[1]{\noindent{{\color{red!80!black}{\sc \bf REV:} \bf   #1}}}
\newcommand{\ahmad}[1]{\noindent{{\color{green!80!black}{\sc \bf AMK:} \bf   #1}}}
\newcommand{\matthias}[1]{\noindent{{\color{olive!80!black}{\sc \bf MM:} \bf   #1}}}
\newcommand{\woong}[1]{\noindent{{\color{blue!80!black}{\sc \bf WS:} \bf   #1}}}
\newcommand{\feiyi}[1]{\noindent{{\color{red!80!black}{\sc \bf FW:} \bf    #1}}}
\newcommand{\naw}[1]{\noindent{{\color{yellow!80!black}{\sc \bf NSS:} \bf   #1}}}
\newcommand{\hao}[1]{\noindent{{\color{yellow!80!black}{\sc \bf Hao:} \bf   #1}}}

%% The actual title
\title{
%Energy Projection and Power Management
%Toward Green Frontier: Savings Projection for Resource-Constrained HPC Application
%Exploring the Frontiers of Software Power Management at System Scale
%Exploring the Frontiers of Energy Efficiency with Software Power Management at System Scale
%Exploring the Frontiers of Energy Efficiency with Software Based Power Management at System Scale
Exploring the Frontiers of Energy Efficiency using Power Management at System Scale
%Quantifying the Frontiers of Software Power Management at System Scale: Telemetry-Informed Energy Saving Projections
%Quantifying Software Power Management at Scale: Telemetry-Informed Energy Saving Projections
%Power Management Projection at Scale: Telemetry-Informed Energy Saving Quantification
%Quantifying Power Management at Scale: Telemetry-Informed Energy Saving Projections

%% Hidden due to the double blind policy
%\thanks{This work was supported by, and used the resources of, the Oak Ridge Leadership Computing Facility, located in the National Center for Computational Sciences at ORNL, which is managed by UT Battelle, LLC for the U.S. DOE (under the contract No. DE-AC05-00OR22725).
%The US government retains and the publisher, by accepting the article for publication, acknowledges that the US government retains a nonexclusive, paid-up, irrevocable, worldwide license to publish or reproduce the published form of this manuscript, or allow others to do so, for US government purposes. DOE will provide public access to these results of federally sponsored research in accordance with the DOE Public Access Plan (http://energy.gov/downloads/doe-public-access-plan).}

%% A temporary subtitle for "Submission number" that facilitates the double blind review
%{\large Submission ID: {pap644}}
%{\large SC24 Submission ID {pap644s1}}
} % \title

% % Specials for author block
% \makeatletter
% \newcommand{\linebreakand}{%
%   \end{@IEEEauthorhalign}
%   \hfill\mbox{}\par
%   \mbox{}\hfill\begin{@IEEEauthorhalign}
% }
% \makeatother

 \author{\IEEEauthorblockN{Ahmad Maroof Karimi}
 \IEEEauthorblockA{\textit{Oak Ridge National Laboratory}\\
 Oak Ridge, TN, USA \\
 karimiahmad@ornl.gov}
 \and
\IEEEauthorblockN{Matthias Maiterth}
\IEEEauthorblockA{\textit{Oak Ridge National Laboratory}\\
Oak Ridge, TN, USA \\
maiterthm@ornl.gov}
%\linebreakand
\and
\IEEEauthorblockN{Woong Shin}
\IEEEauthorblockA{\textit{Oak Ridge National Laboratory}\\
Oak Ridge, TN, USA \\
shinw@ornl.gov}
\and
\IEEEauthorblockN{Naw Safrin Sattar}
\IEEEauthorblockA{\textit{Oak Ridge National Laboratory}\\
Oak Ridge, TN, USA \\
sattarn@ornl.gov}
\and
\IEEEauthorblockN{Hao Lu}
\IEEEauthorblockA{\textit{Oak Ridge National Laboratory}\\
Oak Ridge, TN, USA \\
luh1@ornl.gov}
\and
\IEEEauthorblockN{Feiyi Wang}
\IEEEauthorblockA{\textit{Oak Ridge National Laboratory}\\
Oak Ridge, TN, USA \\
fwang2@ornl.gov}
}

\maketitle

\begin{abstract}
In the face of surging power demands for exascale HPC systems, this work tackles the critical challenge of understanding the impact of software-driven power management techniques like Dynamic Voltage and Frequency Scaling (DVFS) and Power Capping. These techniques have been actively developed over the past few decades. By combining insights from GPU benchmarking to understand application power profiles, we present a telemetry data-driven approach for deriving energy savings projections. This approach has been demonstrably applied to the Frontier supercomputer at scale.
Our findings based on three months of telemetry data indicate that, for certain resource-constrained jobs, significant energy savings (up to $8.5\%$) can be achieved without compromising performance. This translates to a substantial cost reduction, equivalent to $1438$ MWh of energy saved.
The key contribution of this work lies in the methodology for establishing an upper limit for these best-case scenarios and its successful application. This work sheds light on potential energy savings and empowers HPC professionals to optimize the power-performance trade-off within constrained power budgets, not only for the exascale era but also beyond.

\end{abstract}

\IEEEpeerreviewmaketitle


%%%%%%%%%%%%%%%%%%%%%
%% Introduction
\section{Introduction}
\label{sec:intro}

%% Woong - General Context: 100-word
%High-performance computing (HPC) faces significant challenges concerning its energy use and performance growth as we move beyond the Exascale era.
As we transition beyond the exascale era, high-performance computing (HPC) confronts notable challenges related to both its energy consumption and the advancement of its performance capabilities. Power budgets are becoming a major limitation, preventing us from achieving further significant improvements in performance. Simply buying the most power-efficient systems is no longer enough to ensure performance growth. As a result, energy efficiency now depends on how effectively compute resources are utilized during operation. This shift emphasizes the importance of software-driven initiatives and the development of new tools and frameworks to help HPC users and practitioners enhance energy efficiency.

%% Shift of opportunity towards a software driven nature - actually studied very well in the past
In the journey towards exascale, software-driven strategies for enhancing energy efficiency have been acknowledged as effective yet complex avenues towards achieving energy-efficient computing.  By examining various methods of frequency setting~\cite{Cochran2011, Rountree2012, Rong2013,mendes2020exploiting,bharadwaj2023predict} and power capping~\cite{David2010,Rong2017, Patki2019, Ramesh2019, Ghazanfar2022,Krzywaniak2022} the HPC community has recognized that optimizing software can significantly contribute to maximizing performance within constrained power budgets.  These approaches aim to leverage the wealth of information about application behavior and tailor its resource usage and further leverage advanced power management capabilities.

%% The knowledge gap
Despite previous efforts, a significant knowledge gap persists regarding the total impact of software-driven energy efficiency measures across entire HPC data centers. Earlier studies, often based on benchmarks, provide some insights into possible benefits, but mainly due to the difficulty in collecting a comprehensive set of runtime telemetry data, there's a shortage of in-depth, data-driven research examining how optimizations at the application level contribute to overall data center energy savings. Additionally, the quick evolution of computing architectures, particularly towards chiplet-based designs with integrated memory and interconnects, raises questions about the applicability and effectiveness of current strategies in this new landscape. These uncertainties complicate the decision-making process for investing in software technologies that can deliver the required energy efficiency.

%% In this study...
In this study, we tackle the gap in understanding the impact of software-driven energy efficiency on exascale hardware architectures through a data-driven approach. We rely solely on per-component power consumption data, commonly accessible at the system level, to explore this impact on high-performance computing (HPC) systems. 

We begin the study with different power management strategies on benchmark reproducers to understand the power consumption of different resources used. We then apply this analysis to three months of collected per-component power telemetry from the Frontier~\cite{frontier} supercomputer's application jobs. By developing a new method that can estimate key application on-die compute resource usage based on each component's power consumption, we acquire a detailed overview of resource usage across the entire HPC data center. From this comprehensive perspective, we identify several challenges and opportunities by analyzing questions related to software-driven energy efficiency in the post-exascale era.  With this approach, our contributions to the HPC community are as follows:

\subsubsection*{\textbf{Power consumption characterization on contemporary hardware architecture}} Our study provides an updated analysis of the relationship between application-driven on-die resource usage and power consumption in contemporary exascale hardware architectures. Building upon previous research
\cite{David2010,Etinski2012,Rountree2012, Krzywaniak2020}, we specifically examine how advances in power management and the integration of on-die resources like HBM 2e memory and advanced interconnects affect power efficiency. We demonstrate that these architectural innovations significantly alter our understanding of power management's impact in exascale computing environments.

\subsubsection*{\textbf{Power decomposition based telemetry analysis}} We developed a novel power decomposition technique to estimate resource usage in HPC data centers by projecting the benchmark component level characterization to large-scale HPC power consumption data. This approach addresses the challenge of acquiring direct resource consumption telemetry, which can impact system performance and generate vast amounts of telemetry. Our method capitalizes on the detailed insights into application resource usage embedded in power consumption data, available through out-of-band channels in contemporary HPC systems \cite{bartolini2018davide, Thaler:2020, openbmc_event, shin2021revealing}.

\subsubsection*{\textbf{Holistic insights and recommendations on the impact of software-driven energy efficiency}} Through our study, which includes estimating resource usage at the data center level and analyzing power consumption, we have extensively examined the impact of software-defined energy efficiency. This investigation uncovers numerous challenges and opportunities within this approach emerging due to the non-trivial shift in both hardware architecture and the workloads we see on exascale systems. Here, we offer key findings to the HPC community to maximize software-driven energy efficiency.

%Through our study, we reveal the diminishing opportunities in the DVFS based power management functionality and the increasing importance of middleware and runtime level optimizations that can help applications increase utilization of on-die resources.
%(TODO: modify based on a quick summary of key insight we identify from our study - no need to outline the next sections)

% The goal of this paper is to project 
% application power based on potential power management controls used. The anticipated outcome is to understand the impact power management can have on the data center.
% For this, we decompose GPU power decomposition, which allows us to map GPU power profiles to GPU operation profiles.
% %
% The main contributions of this paper are:
% \begin{itemize}
% \item We design and develop a methodology for GPU power projection
% \item Analysis  of GPU power decomposition and perform analysis on mapping GPU power with GPU operations
% \item Recommendations based on power projection analysis
% \end{itemize}

% Sections~\ref{sec:background} covers

%%%%%%%%%%%%%%%%%%%%%
%% Background
\section{Background and Related Work}
\label{sec:background}
The need for accurate power prediction is essential for large data centers.
To assess the potential for improvements in energy efficiency we need to put previous work and current technology in context.

%%%%%%%%%%%%%%%%%%%%%%%%%%%%%%%%%%%%%%%
%% Subsection 1 - Architecture
\subsection{Power Implications of Heavy/Fat CPU+GPU Heterogeneous Node Architecture Towards Exascale}
%\subsection{Power Hungry GPUs in a Heavy/Fat CPU+GPU Heterogeneous Node Architecture Towards Exascale}

In the quest for exascale computing, the Oak Ridge Leadership Computing Facility (OLCF) has progressively evolved through three major supercomputing systems: Titan~\cite{titan}, Summit~\cite{summit}, and Frontier~\cite{frontier}. The architectural journey commenced with Titan, which introduced a CPU+GPU-based heterogeneous node design. This approach was further refined with Summit, serving as a stepping stone toward achieving exascale capabilities with the Frontier supercomputer.

A prominent feature shared among these systems is their heavy/fat node architecture that increasingly favored GPUs over CPUs—from a 1:1 CPU to GPU ratio in Titan to 1:3 in Summit, and ultimately 1:4~\footnote{1:8 considering the two Graphic Compute Dies (GCD)} in Frontier~\cite{atchley2023frontier}. Additionally, while Titan initially relied on air cooling, the subsequent need to manage the heat from power-intensive GPUs led Summit and Frontier to adopt direct liquid cooling techniques. These later systems enhanced energy efficiency by utilizing medium or high-temperature water in their cooling loops, facilitating evaporative cooling methods~\cite{shin2021revealing}.

Beginning with Summit, the adoption of \gls{HBM2e} for GPUs became a standard to circumvent data movement bottlenecks. This shift not only catered to the increasing computational demands with GPUs as the primary compute drivers within these systems. Given their high-parallelism \gls{SIMD} capability coupled with \gls{HBM2e} memory, GPUs emerged as the most significant power consumers within the HPC user facility.

%%%%%%%%%%%%%%%%%%%%%%%%%%%%%%%%%%%%%%%
%% Subsection 2
%\subsection{Energy saving opportunities of dynamic power management with contemporary hardware}
\subsection{Energy saving opportunities using dynamic power management on contemporary hardware}

% \woong{prior studies:
% \begin{enumerate}
% \item Software Defined EE is not evaluated at contemporary harware at exascale - large scale. 
% \item -- We need proof for these.
% \end{enumerate}
% }

Dynamic power management in HPC has been a well-studied area for more than a decade since the race towards exascale \cite{Cochran2011, Rountree2012, Rong2013,David2010,Hackenberg2015,Rong2017, Patki2019, Ramesh2019, Ghazanfar2022,Krzywaniak2022}. Yet, the evolution toward chiplet-based architectures and 3D stacked high-bandwidth memory~\cite{Loh2023,Kogge2022} necessitates a fresh look at dynamic power management strategies. Previous studies, mainly from the 2010s, focused on \gls{DVFS} to gauge its effect on application performance \cite{Cochran2011, Rountree2012, Rong2013,David2010}. These investigations highlighted \gls{DVFS} as an effective method for reducing dynamic energy consumption, especially in memory-intensive applications that encounter bottlenecks due to off-chip data movement \cite{Wang2020,Cesarini2020}. Some of which investigated \gls{DVFS} on \gls{SIMD}-based architectures.

%Dynamic power management in HPC is a well studied area for more than a decade since the race towards exascale \cite{Cochran2011, Rountree2012, Rong2013,David2010,Rong2017, Patki2019, Ramesh2019, Ghazanfar2022,Krzywaniak2022}. Yet, the evolution toward chiplet-based architectures and 3D stacked high-bandwidth memory~\cite{Loh2023,Kogge2022} necessitates a fresh look at dynamic power management strategies. Previous studies, mainly from the 2010s, focused on \gls{DVFS} to gauge its effect on application performance \cite{Cochran2011, Rountree2012, Rong2013,David2010}. These investigations highlighted \gls{DVFS} as an effective method for reducing dynamic energy consumption, especially in memory-intensive applications that encounter bottlenecks due to off-chip data movement \cite{Wang2020,Cesarini2020}. Some of which investigated \gls{DVFS} on \gls{SIMD}-based architectures.

Recent explorations into the dynamic power management capabilities of contemporary GPUs, such as \cite{Krzywaniak2019,White2022}, strive to update our understanding by reassessing the impact of power management on systems and application performance. These studies confirm that it remains a viable strategy for conserving energy in modern GPU environments. Yet, they are often too narrowly focused on specific applications, lacking a broader generalization about energy-saving opportunities across varying workloads.

%%%%%%%%%%%%%%%%%%%%%%%%%%%%%%%%%%%%%%%
%% Subsection 3
\subsection{Dynamic power management in the HPC data center}

%\woong{REVISIT and update after methodology is complete}

% \woong{Prior studies:
% \begin{enumerate}
% \item Knowledge gap -- No one answered the impact of power management at the Datacenter level
% \item -- We need proof for these.
% \item Usually people do not have the telemetry data, therefore its hard to do the technology / knowledge transfer.
% \item These studies are lacking either: 
% \begin{itemize}
% \item Granularity
% \item Scope
% \item Duration
% \end{itemize}
% \item telemetry data is missing
% \item byproduct of modality study: Modality and component level monitoring or telemetry is not feasible (we are at 15 seconds, others are at 1m or 15m this is not widely available.)
% \item we want to understand our center. 
% \end{enumerate}
% }

The HPC community has long recognized the potential benefits of dynamic power management for enhancing energy efficiency within data centers \cite{FacebookDynamo2016,ALMEIDA2018}. Despite the challenges in generalizing the impact due to diverse architectures and applications across HPC systems, previous studies have shed light on its significant potential. These investigations range from benchmark~\cite{Leon2016} and modeling-driven studies~\cite{Chadha2021,Elisseev2018} to those leveraging always-on telemetry data \cite{Wallace2016}, each contributing valuable insights yet revealing gaps in our comprehensive understanding. Specifically, benchmark-driven approaches often fail to fully encapsulate the entirety of the system \cite{Leon2016}, while telemetry data-driven studies \cite{ALMEIDA2018} struggle with collecting and managing extensive datasets needed for depth analysis. Modeling attempts, though promising, have traditionally grappled with a lack of detailed data \cite{Wallace2016}. Recognizing these limitations, our approach aims to synthesize these methodologies by combining benchmarks, rich telemetry data, and novel modeling techniques.  Compared to the reports from CPU-based HPC systems that have deployed dynamic power management as one of the major energy efficiency strategies, we aim to shed light on the impact in-priori with potential insights on a GPU-based system.

\section{A Hybrid Methodology}
\label{sec:methodology}
%\ahmad{}%Overview of methodology 50-words }
This work aims to analyze Frontier’s GPU operations
and their power utilization in real-world HPC applications
by leveraging the understanding from different benchmark runs. We study the impact of applying frequency and power management on GPUs and its effect on performance and power
utilization. To analyze the impact of software-driven energy efficiency on exascale hardware architectures. We break our methodology into three main parts: 
Collecting power data on the Frontier system, evaluating the power consumption of benchmark reproducers, and mapping consumption patterns with HPC Applications for data-driven analysis.

%% Woong: I think this one should go upwards into the "background" section

%\subsection{Telemetry Data Collection and Usage}
\subsection{Utilizing Power Telemetry Data}

\begin{table}
  \scriptsize
  %\footnotesize  
  \centering
  \caption{Frontier system's summary}
  \label{tab:frontier_specs}
  %% Footnotesize if a much smaller font is needed
  %\footnotesize
  \begin{tabular}{lc}
    
    \toprule
     \multicolumn{2}{c}{Frontier System}\\ 
    \midrule
     % Name & colB  \\
    %\midrule
      Compute node & 9408 \\
      Peak performance & 1.9 EF \\
      Peak power & 29 MW \\
      GPU memory (HBM) & 4.6 PB \\
      CPU memory (DDR4) & 4.6 PB \\  %\midrule
      %\multicolumn{2}{c}{Compute node}\\
      %\midrule
        Each Compute node & 4 AMD MI250X \\
        Each GPU & 2 GCD \\
        Each GCD & 64 GB HBM2E \\
        GCD max power & 560 W \\
        GCD max frequency & 1700 MHz\\
        HBM bandwidth & 1.6 GB/s\\
        %GCD peak performance & 23.95 TFLOP/s\\
  \bottomrule

\end{tabular}
\end{table}

Telemetry collection for Frontier is an ongoing effort~\cite{doubleblind_telemetry}, which we use to assess improvements in energy efficiency.

\begin{figure}[t]
  \centering
  \includegraphics[width=\linewidth]{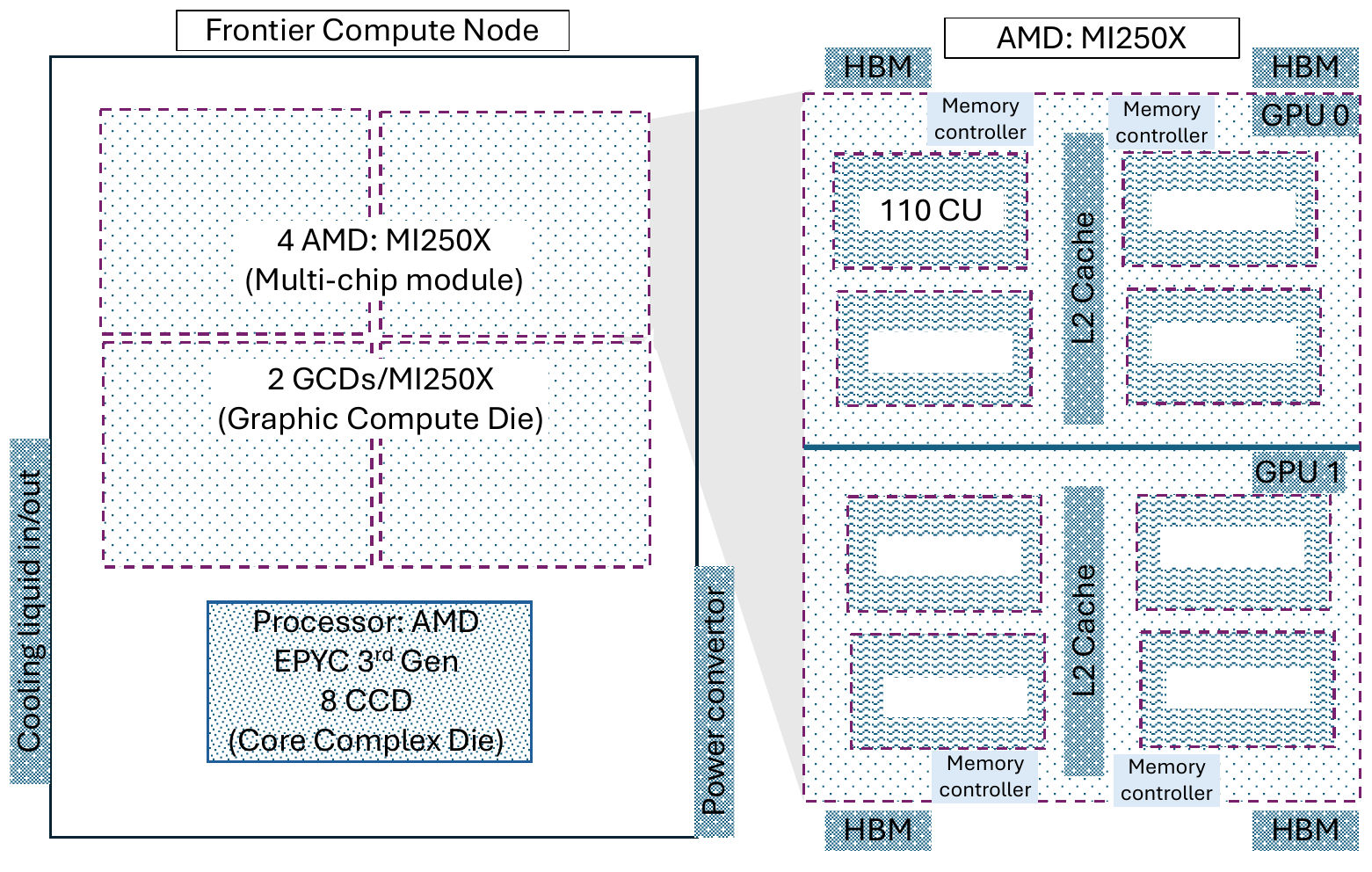}
  \caption{Schematic representation of Frontier compute node and MI250X multi-chip GPU}
  %\Description{Description of the image}
  \label{figs:CNArch}
\end{figure}

A summary of the Frontier system specification is shown in Table~\ref{tab:frontier_specs} and an abstract representation of a Frontier compute node along with its 4 GPUs is illustrated in Figure~\ref{figs:CNArch}. 
Every compute node in Frontier has 4 AMD: MI250X GPUs. The GPUs are built on advanced packaging technology, which enables 2 Graphic Compute Dies (GCDs). Each GCD has access to 64 GB High Bandwidth Memory (HBM) and FP64 peak performance of 23.9 TFLOPS/s. To the end-users, each GCD appears as a GPU and can be accordingly used for workload submission.

We collect 3 months of Frontier's real-world applications' telemetry data for analysis and energy projections.  
We then perform benchmark simulations to map real-world telemetry to get insights into workload behavior. 
%The benchmarks include GPU roofline-model and GPU micro-benchmarks to mimic kernel operations in applications exhibiting different modes of operations.  
We use the telemetry logs to analyze the application power utilization and to identify the modes of operations in real-world applications and how the frequency and power capping will impact them based on the results from benchmark models.
%Based on the combined analysis of benchmark runs and HPC applications, we project energy-saving opportunities.

%% Template for a page-wide figure

\begin{figure}[t]
  \centering
  \includegraphics[width=0.9\linewidth]{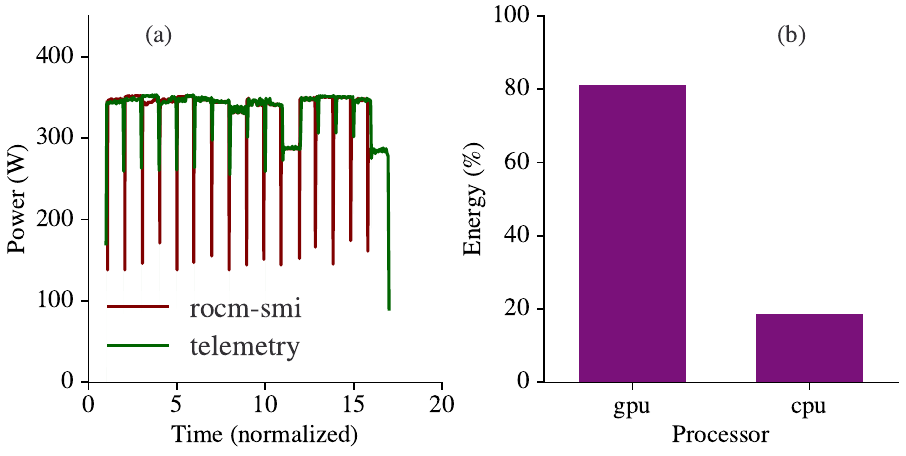}
    \caption{Plot (a) compares telemetry data and ROCm SMI data for a sample Frontier application run. The histogram in the plot (b) shows the histogram of GPU and CPU energy utilization on the Frontier system.}
  \label{plots:rocmsmi-tele}
\end{figure}

Modern exascale HPC systems like Frontier use an extreme amount of power for their day-to-day operations. The performance of these large-scale machines is primarily powered by GPU resources as shown in Figure~\ref{plots:rocmsmi-tele}(b). In this study, the focus of our energy analysis and saving projections is geared towards understanding GPU energy utilization patterns and their saving projection opportunities. For analysis we use the HPC compute node telemetry data and the data from the Jobs scheduler log for metadata information. We also show in Figure~\ref{plots:rocmsmi-tele}(a) that the telemetry data is comparable to the data derived from the ROCm SMI library (a tool to monitor GPU utilization) \cite{rocmsmi} which we used during benchmark analysis. 
A summary of data is presented in Table~\ref{tab:telemetry_data}.  
 
\paragraph{Telemetry Data Description}
%Refer to SC'21 paper and briefly describe it here.

%% Wide table
\begin{comment}

\begin{table*}
  \centering
  \caption{Telemetry Dataset summary}
  \label{tab:telemetry_data}
  %% Footnotesize if a much smaller font is needed
  \footnotesize
  \begin{tabular}{ccccl}
    \toprule
    id & Name & Resolution & Rows & Description\\
    \midrule
    \texttt{(a)} & Power telemetry data  & 15 sec. & &  Out-of-band data collected from each Frontier node\\
    \texttt{(b)}& Job scheduler log& per-job& & Metadata for each job, \texttt{num\_nodes}, \texttt{project\_id},\texttt{begin\_time} and \texttt{end\_time}\\
    \texttt{(c)}& Per-node Scheduler data& per-node-per-job& &Meta-data information about jobs running on each compute node \\
    \bottomrule
  \end{tabular}
\end{table*}

\end{comment}

\begin{table}
  \centering
  \caption{Telemetry Dataset summary}
  \label{tab:telemetry_data}
  %% Footnotesize if a much smaller font is needed
  \footnotesize
  \begin{tabular}{cm{1.7cm}m{1.5cm}m{3.5cm}}
    \toprule
    id & Name & Resolution & Description\\
    \midrule
    \texttt{(a)} & Power telemetry data  & 15 sec. &   Out-of-band data collected from each Frontier node\\
    \texttt{(b)}& Job scheduler log& per-job&  Metadata for each job, \texttt{num\_nodes}, \texttt{project\_id}, \texttt{begin\_time} and \texttt{end\_time}\\
    
    \texttt{(c)}& Per-node Scheduler data& per-node-per-job& Metadata information about jobs running on each compute node \\
    \bottomrule
  \end{tabular}
\end{table}

%For this paper we heavily rely on the telemetry pipeline in production at OLCF. 
The infrastructure has been introduced even before the Frontier system and is evolving alongside the production systems, continuously upgraded with new installations. 
The setup is described in the architecture section of the work by~\cite{doubleblind_telemetry}.
%Shin et.\@al.\@~\ref{shin2021revealing}.
The telemetry contains node power input, with explicit GPU and CPU power utilization. These features are captured by sensors attached to the HPC's compute nodes. %These are system-level information and are collected independent of whether the jobs are running on compute nodes or not. 
The logs are captured at a frequency of 2-second intervals and are aggregated in the pre-processing state to make it 15-second intervals. 
The GPU-level power utilization data enables us to study the behavior of GPUs under HPC workloads that exert diverse severity of stresses on bandwidth, memory, or computing resources. However, telemetry data lacks metadata information on workloads, projects, and other fields, which is also crucial. 

\paragraph{Job-scheduler Log Description}
The rich telemetry data is necessary to comprehensively understand the job's power utilization behavior. 
However, the application level information such as \texttt{job\_ids} and \texttt{science\_domains} are collected from scheduler logs generated by SLURM. Job-scheduler logs provide us with information about jobs that help us identify the start- and end-time of a job and the compute nodes on which the job was executed. Joining job-scheduler logs and telemetry data is essential for analysis at the jobs and science domain level. 

%To get science domain and application name of jobs we refer to jobs log, Shin et.\@al.\@~\ref{arxiv_paper}.

%%%%%%%%%%%%%%%%%%%%%%%%%
%% The benchmark

\subsection{Characterizing Power Consumption using Benchmarks}
\label{subsec:benchmarks}
%As systems like Frontier are designed for the large-scale high-performance codes, two approaches can be taken to analyze optimized codes running on the system. This can be achieved either by using proxy applications or synthetic workloads.
Systems such as Frontier are designed for large-scale high-performance codes. There are two methods available for analyzing optimized codes running on the system. This can be accomplished either through the utilization of proxy applications or synthetic workloads.

Proxy applications represent a kernel of a full application workload without the complexity of the entire application. Synthetic workloads, on the other hand, aim to stress specific sub-systems, such as the \gls{ALU}, and the Memory Subsystem, among others, to understand the achievable performance of the system based on its specifications.
To understand how energy is utilized by the system and how to improve it, it is crucial to understand the interplay between power consumption and performance or time to solution. We executed various benchmarks to observe and analyze the power and performance characteristics of GPUs by varying frequencies and power caps. Additionally, we varied the size of the input data and arithmetic intensity to analyze the characteristics of different resource utilization. In the following subsection, we provide details of the benchmarks we used to conduct the study.

\paragraph{Roofline and \gls{VAI} Benchmark}
The roofline model is a useful tool to understand the maximum attainable performance on a system ~\cite{Williams:EECS-2008-164}.
It provides reference points on the achievable performance.
%A typical assessment of maximum achievable performance is the roofline model~\cite{Williams-Roofline2009}.
The usual graph resulting from this is a roofline plot, showing maximum achievable performance, considering the peak \gls{FLOPS/s} and memory bandwidth, given arithmetic intensity in operations per byte.
Tools such as the Empirical Roofline Model~\cite{Yang2018}, allow the assessment of a system in a relatively portable way of measuring hardware performance.
%The result is typically graphed in a plot where the y-axis shows \gls{FLOPS/s} and the x-axis is the arithmetic intensity in FLOPS/Byte.
For this \gls{ERT}~\cite{ERTonline} uses different \gls{FMA} micro-kernels for memory, as well as an \gls{GPP} kernel for the compute kernel~\cite{Yang2018}.
To measure power accurately, the runtime has to be longer than in typical~\gls{ERT} runs. Additionally, to understand power behavior along the roofline, we want to sweep along the performance roof while taking power measurements. Therefore we designed a synthetic benchmark to directly trace the roofline, allowing us to adjust runtime to get accurate power measurements.

To perform these measurements we implement Algorithm~\ref{alg:VAI}. The algorithm takes arithmetic intensity as the input parameter.
%, which can be called given specific arithmetic intensity. 
With this, we can easily measure and reproduce the performance lines observed in the roofline model by changing arithmetic intensities and measuring runtime and sustained power. 
%This implementation allows us to arbitrarily extend the runtime to be able to observe the steady state power behavior of the kernel while preserving the characteristics in arithmetic intensity with known parameters and data size.
This implementation allows us to extend the runtime of the kernel. This extension ensures we can observe its steady state power behavior while maintaining the desired characteristics of arithmetic intensity using predetermined parameters and data sizes.
To generate the roofline plot, we count the number of operations performed and the number of bytes transferred and compare different runs.

\begin{algorithm}
\footnotesize
\caption{Variable Arithmetic Intensity: VAI Benchmark}\label{alg:VAI}
\SetAlgoLined
\SetKwData{GlobalWIs}{\texttt{globalWIs}}
\SetKwData{Repeat}{\texttt{REPEAT}}
\SetKwData{LoopSize}{\texttt{LOOPSIZE}}
\KwData{%
\GlobalWIs: number of work-items\;
\qquad\:\:\,\,\Repeat: number of repetitions\;
\qquad\:\:\,\,\LoopSize: number of repetitions to achieve
\qquad\:\:\quad$2*\LoopSize$ operations per 4 Loads/Stores\;
\qquad\:\:\,\,$a \gets [globalWIs]$, where $a[i] \gets 1.3$\;
\qquad\:\:\,\,$b \gets [globalWIs]$, where $b[i] \gets i$\;
\qquad\:\:\,\,$c \gets [globalWIs]$, where $c[i] \gets 1.3$\;
}
\KwResult{Array $c$}
\For{$j \gets 0$ \KwTo \Repeat}{
    \#pragma omp target teams distribute parallel for simd\\
    \For{$i \gets 0$ \KwTo \GlobalWIs}{
          $x \gets a[i]$ \tcp{Read 1}
          $y \gets b[i]$ \tcp{Read 2}
          $z \gets c[i]$ \tcp{Read 3}
          \#pragma unroll(\LoopSize)\\
          \For{$lc \gets 0$ \KwTo \LoopSize}{
            $z \gets x \times y + z$ \tcp{2 ops}
          }
          $c[i] \gets z$ \tcp{Write 1}
    }
}
\end{algorithm}

The algorithm allocates 3 arrays, a, b, and c, with a large enough number of global work-items $\texttt{globalWIs}$ to fill up the GPU memory. Additionally, the constant $\texttt{REPEAT}$ is set so that the code runs at least for 20 seconds. 
The for loop on line 3 now repeatedly iterates over all elements and performs the following:
Read the elements of the array (3 Reads), perform $2 * \texttt{LOOPSIZE}$ operations (line 7--10), and write back the result (1 Write). 
Given the datatype, we can calculate the arithmetic intensity from these operations.
The runtime and power are measured, while the size of globalWIs is known, allowing us to calculate achieved FLOPS/s and the memory bandwidth in GByte/s. 
For double we achieve as low as $1/16$ arithmetic intensity, For arithmetic intensity of $0$ the lines 7--11 are replaced by $\texttt{c[i]}\xleftarrow{}\texttt{b[i]}$.
We achieve $>90\%$ performance on MI250X for this simple algorithm, with contiguous memory access and \texttt{SIMD} parallelism while the OpenMP runtime takes care of saturating the GCD. To run this on all GCDs of the MI250X we run this in an embarrassingly parallel fashion using MPI, where each GCD operates on a separate copy of its own data, as we want to trigger specific performance to be able to measure the behavior rather than doing useful work. 
With this we can mimic the high peak performance a typical user should achieve, given a simple algorithm without excessive optimization, implemented as a simple embarrassingly parallel kernel.

To extend the runtime for steady state measurement we increase the $\texttt{REPEAT}$ timer so that we observe stable power measurements.
%With this implementation for a roofline benchmark we can measure power and runtime, while setting number of workitems and arithmetic intensity.  
This VAI benchmark allows us to show the limits of maximum utilization of a system in a roofline fashion, while able to associate sustained power to performance characteristics. 

% %% Template for a page-wide figure
% \begin{figure}[t]
%   \centering
%   \includegraphics[width=\linewidth]{figs/template_single.pdf}
%   \caption{Figure for describing benchmark simulation workflow}
%   %\Description{Description of the image}
%   \label{figs:template_single}
% \end{figure}

\begin{figure}[t]
  \centering
  \includegraphics[width=.8\linewidth]{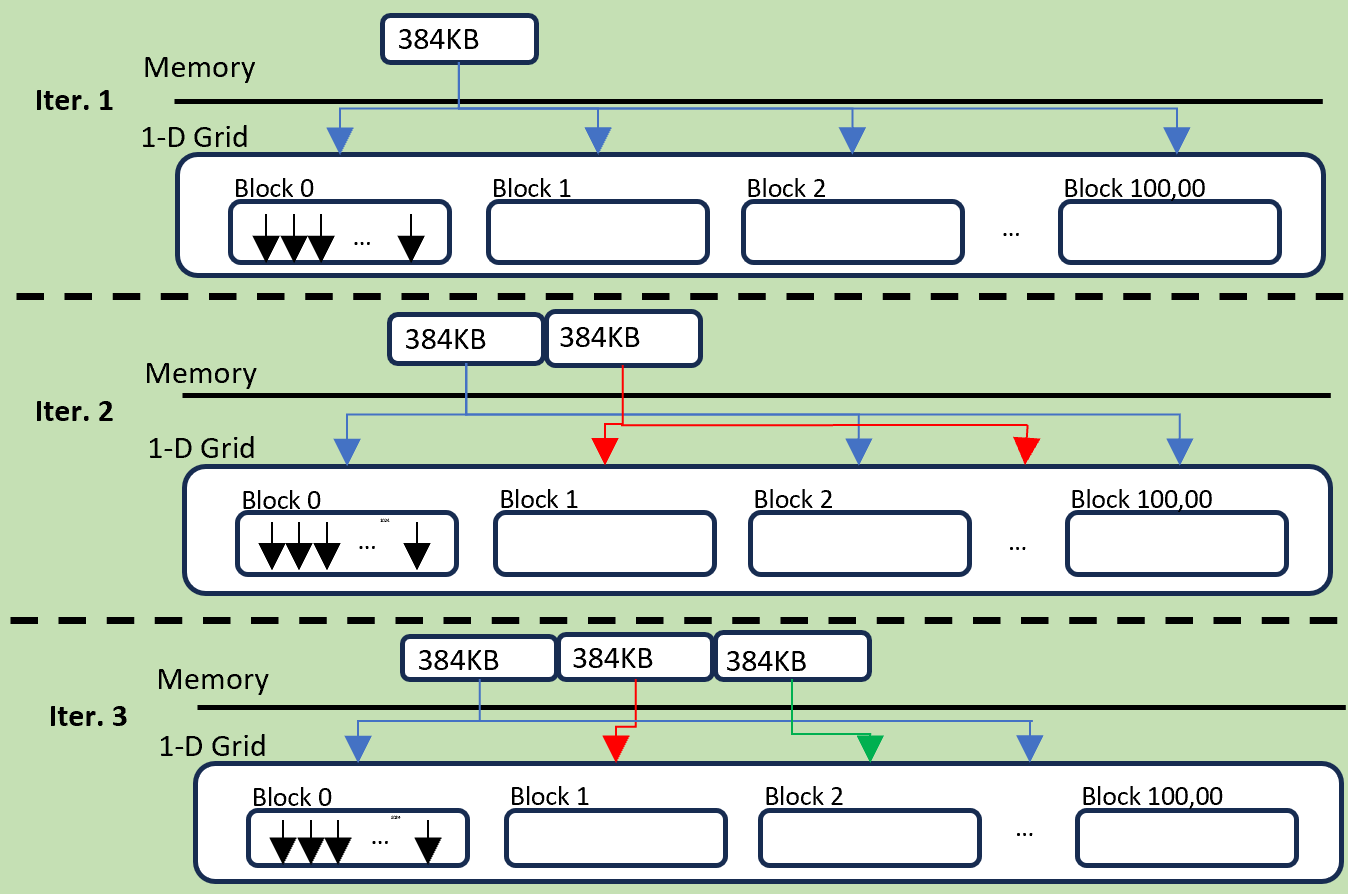}
  \caption{GPU benches L2-cache memory access pattern}
  %\Description{Description of the image}
  \label{figs:l2-cache}
\end{figure}

\paragraph{Benchmark for GPU Memory Characterization}
To gain insights into GPU power consumption during memory-insensitive executions, we consulted published work and conducted tests using a modified version of the L2-cache benchmark within GPU-Benches~\cite{gpu-benches,Ernst2021}. This benchmark is tailored to gauge the maximum bandwidth of the L2-cache and main memory across various memory load sizes. The benchmark starts with a single memory chunk of $384KB$ and incrementally increases the chunk size. Loading operations initiate a kernel comprising $100,000$ blocks, each with $1,024$ threads, processing a chunk equal to blocks\_id \% number of memory chunks. This kernel effectively saturates the bandwidth by repeatedly loading the same chunk to different blocks, detailed in Figure~\ref{figs:l2-cache}. The premise behind this exploration is rooted in the notion that if GPU performance is constrained by memory bandwidth, slowing down the GPU while it awaits memory access could yield energy savings.

%\paragraph{Verification of Memory Characteristics using a Graph Benchmark Application}
\paragraph{Verification of Memory Characteristics using a Real HPC Graph Application}
\label{method:graph}
%\naw{Graph application description and how it is being used to understand gpu utilization pattern}
To gain confidence in the power characteristics of the GPU memory cache benchmark, we use a workload that is sensitive to memory latency and memory bandwidth characteristics.
GPU-based Louvain community detection is a valuable workload to represent a real application due to its mixed memory and compute-intensive characteristics. Community detection is a fundamental task in graph mining \cite{blondel,lu2015parallel}, encompassing domains like social networks, web networks, road networks, and many more. The Louvain algorithm \cite{newman2004finding} involves a combination of compute-intensive modularity calculations and data movement between CPU and GPU. This mix reflects the nature of many graph algorithms, making it a well-rounded application for the use case study. Larger and denser graphs generally demand more processing power and memory bandwidth from the GPU, resulting in higher utilization and power draw. Whereas sparse graphs might underutilize the GPU, leading to lower power consumption. But that is not always the case considering thread divergence and other bottlenecks arose for the irregular graph applications. To get an actual understanding of the GPU power usage, these experiments provide valuable insights into the GPU's capabilities for handling complex graph workloads, making it a beneficial addition to the overall benchmarking analysis. In this GPU-based Louvain community detection application, the input graphs are processed in a Compressed Sparse Row (CSR) format, for more regular memory access. The GPU compute workload distribution is done among the threads based on the degree distribution of the vertices in the network to efficiently handle both power-law (scale-free) networks and bounded networks. The overall workload can be represented by the number of edges in the network. A detailed description of this GPU-based graph algorithm can be found in \cite{doubleblind_gpu_louvain}. For this analysis, we choose input networks \cite{snap} with variations in edges [$3K-8M$] and degree distributions [$d_{max}:9 \text{ to } 343$\text{, }$d_{avg}:2\text{ to }23$] of the networks. Here $d_{max}$ and $d_{avg}$ denote the maximum degree and average degree.

\subsection{Data-driven Analysis for Modal Decomposition}
\label{subsec:method_modal_decomposition}
Analysis of the Frontier's HPC workloads is essential for understanding the behavior of GPU power utilization in real-world applications. We combine the telemetry power data and job-scheduler log to analyze the power data at job-level power. The job-level power data has GPU power utilization at a 15-second frequency from each GPU on every node on which the workload was executed. Job-level GPU power utilization enables us to gain deeper insights into applications' compute operations and memory utilization. Since the power data is an instantaneous value and not an aggregated value, it gives a fair sweep of the characteristics of workloads. 

We analyze the distribution of GPU power to identify the most prevalent values of utilized power and the corresponding GPU operations. We perform the analysis on a system-wide scale and at the science domain level. At a system-wide scale, we see the combined pattern of all jobs, whereas, in the science domain, jobs are much more similar and exhibit a similar GPU power utilization pattern. Distinctive patterns help us better understand the nature of GPU computing and memory utilization.  
Combining the data of real-workload utilization and mapping it with the roofline and GPU microbenchmark metrics on the same GPU model allows us to identify the operations HPC workloads perform for a given range of power utilization. However, it is not possible to disaggregate all the GPU operations based only on the power values. We intend to have a collection of operations with similar power values in one group that are significantly different from other groups. For each group or mode, we illustrate their potential range of operations based on the analysis of benchmark applications and roofline models.

% \ahmad{}% Workload Classification 150-words: Classification into Compute intentesive, Non-compute, and Mixed workloads }
% Prior to identifying modes of operations, we classify the HPC jobs into three classes, i.e., compute-intensive, non-compute-intensive, and mixed-intensity jobs. The classification of workloads based on the compute intensity enables us to visualize the modes of operation in different classes. 

\subsection{Mapping the observed benchmark behavior to the identified modes for the HPC applications on the full system scale}
\label{subsec:method_mapping_the_modes}
An important contribution of our work is to leverage the benchmarks to understand the patterns exhibited by real applications.
The goal is to use these for a projection of an upper limit on potential energy savings, the best-case scenario.
For this task, we classify the GPU operations into four power ranges: ideal, non-compute resource bound, compute bound, and boosted frequency range. Mapping roofline and benchmark results with the power values from the observed data in actual HPC workloads or applications is essential for understanding GPU modes of operations and projecting energy savings estimates. 

GPUs execute hundreds of kernels or mathematical operations; individually analyzing the power behavior of all GPU kernel operations is not required because, from the power utilization perspective, many operations use similar power values, and all operations can be combined into a few groups. 
Another usefulness of grouping the GPU power values based on modes of operations is that it makes it feasible to examine the low-level power and energy utilization patterns of real-world HPC workloads. 
The energy utilization pattern provides energy projection and estimation of savings opportunities. 

With this methodology in mind and the telemetry data available, we perform the GPU benchmark analysis in Section~\ref{sec:gpu_characterization_analysis}. It is followed by the modal decomposition and power projection in Section~\ref{sec:Projection} to obtain upper bounds for the potential of energy savings using the available power management techniques. 
%With this methodology in mind, and the telemetry data available, we perform the GPU benchmark analysis in the following Sections. Section~\ref{sec:gpu_characterization_analysis}, followed by the modal decomposition and power projection in Section~\ref{sec:Projection} to obtain upper bounds for the potential of energy savings using the techniques available on the system.

%%%%%%%%%%%%%%%%%%%%%
%% Analysis and Experiment
\section{
%GPU Characterization Benchmark
Characterizing Power Behavior using GPU Benchmarks
}
\label{sec:gpu_characterization_analysis}
%The objective of the GPU power consumption characterization benchmark is to execute different scenarios and to observe GPU performance under different operating conditions. We analyze the power and performance characteristics of GPUs under varying frequencies and power caps. We also vary the size of the input data and arithmetic intensity to analyze the characteristics of GPU power utilization. We use the benchmark characteristics to understand the behavior of Frontier applications based on the distribution of power utilization.

In this section, we analyze the power data collected using the benchmark described in Section~\ref{subsec:benchmarks}. We analyze this power consumption data to establish a characterization basis for real applications.

\subsection{Power Characteristics of the Roofline on MI250X}
As discussed in Section~\ref{subsec:benchmarks}-a, the roofline model shows the theoretical peak, according to peak FLOPS/s and peak memory bandwidth, for different arithmetic intensities in operations per byte. 
Using the \gls{VAI} Benchmark we probe the peak achievable performance and understand the power consumption behavior. We aim to comprehend the power consumption across different arithmetic intensities and understand the impact of power management on each arithmetic intensity.

\begin{figure}
    \centering
    \includegraphics[width=0.48\columnwidth]{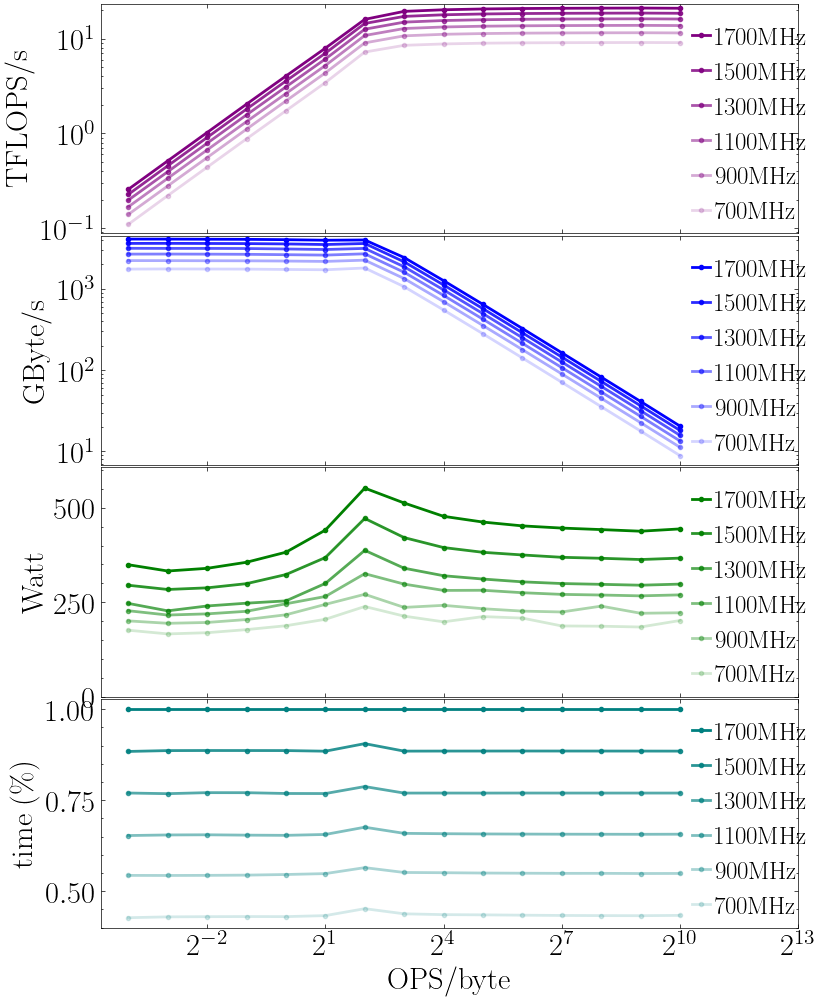}
    ~
    %\centering
    \includegraphics[width=0.48\columnwidth]{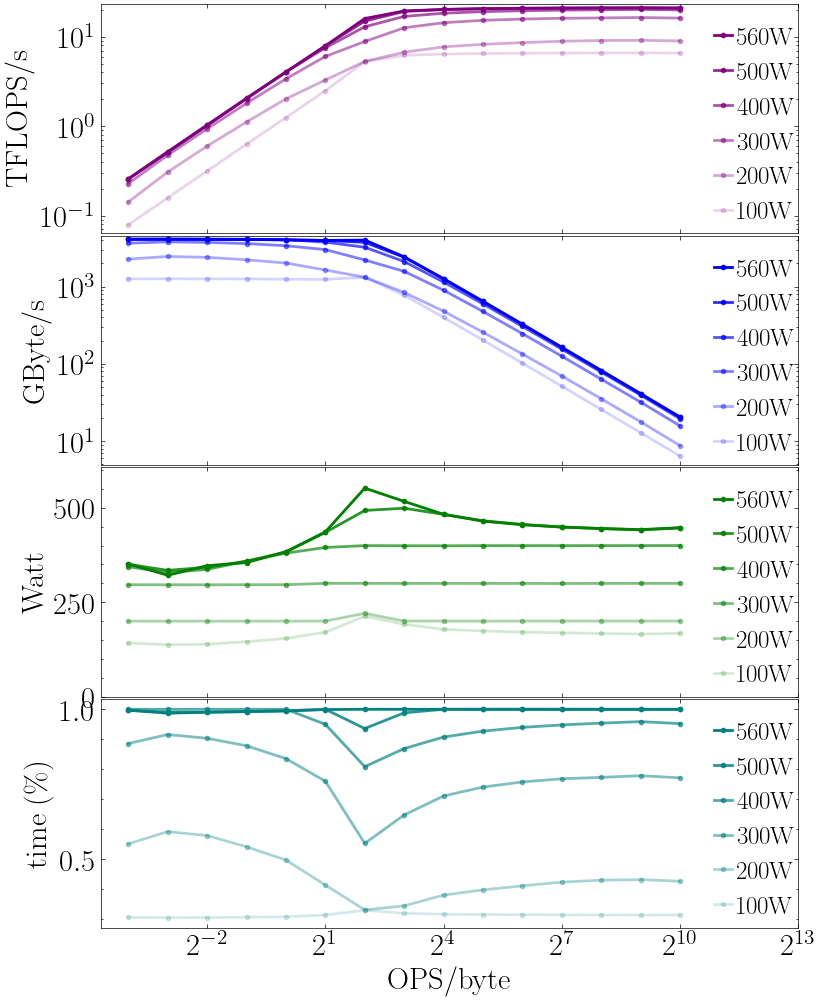}
    \caption{Roofline Plot showings: Left: Fixed Frequency, Right: Power Cap. Top to bottom: a) TFLOPS/s, b) GByte/s, c) Power Consumption, d) normalized time to solution, with different power limits for a single GPU, while running all tiles of an MI250X. The x-axis is the arithmetic intensity by operations per byte.}
    \label{fig:roofline}
\end{figure}
We executed Algorithm~\ref{alg:VAI} for each arithmetic intensity, experimenting with frequency ranges from $1700$ to $700$ MHz, and with power caps set from $560$W to $100$W, shown in the left and right column of Figure~\ref{fig:roofline}. The runtime is always normalized to $1$ for each arithmetic intensity to evaluate the impact of power management mechanisms for a specific arithmetic intensity.

The left column of Figure~\ref{fig:roofline} illustrates the frequency dependence of both the memory-bound part, with arithmetic intensity below four ($2^2$), and the FLOPS/s-bound part, beyond an arithmetic intensity of 4. We observed the maximum power consumption of the GPU is 540W, which is reached only at the maximum frequency and an arithmetic intensity of 4. The consumption ranges from 380W at an arithmetic intensity of 1/16 to 540W at an arithmetic intensity of 4, and then decreases to 420W. Adjusting the frequency downward reduces consumption proportionally, but with a corresponding impact on performance.

%The left column of Figure~\ref{fig:roofline} shows frequency dependence of both the memory bound part with arithmetic intensity below four ($2^2$), as well as for the FLOPS/s bound part beyond arithmetic intensity of 4. The runtime is proportional to the frequency selected. The interesting part is to consider the power consumption of a GPU: The maximum power consumption of the GPU is 540\,W. At Maximum frequency this is consumed only at an arithmetic intensity of 4.The consumption sweeps from 380\,W at arithmetic intensity of 1/16 to 540 at arithmetic intensity of 4 down to 420\,W. Adjusting the frequency down, the consumption is lowered proportionally, but similarly with a proportional impact to performance.

Notable takeaways include: Read/write-heavy applications show significant power consumption. FLOPS-intensive applications exhibit slightly higher peak power consumption but do not reach peak power. Only when stressing both the memory subsystem and the ALUs is the \gls{TDP} reached. Since memory accesses are contiguous, both memory and FLOPS-bound parts are affected by frequency throttling similarly on the given architecture.

\begin{figure}[t]
    \centering
    \includegraphics[width=1.000\columnwidth]{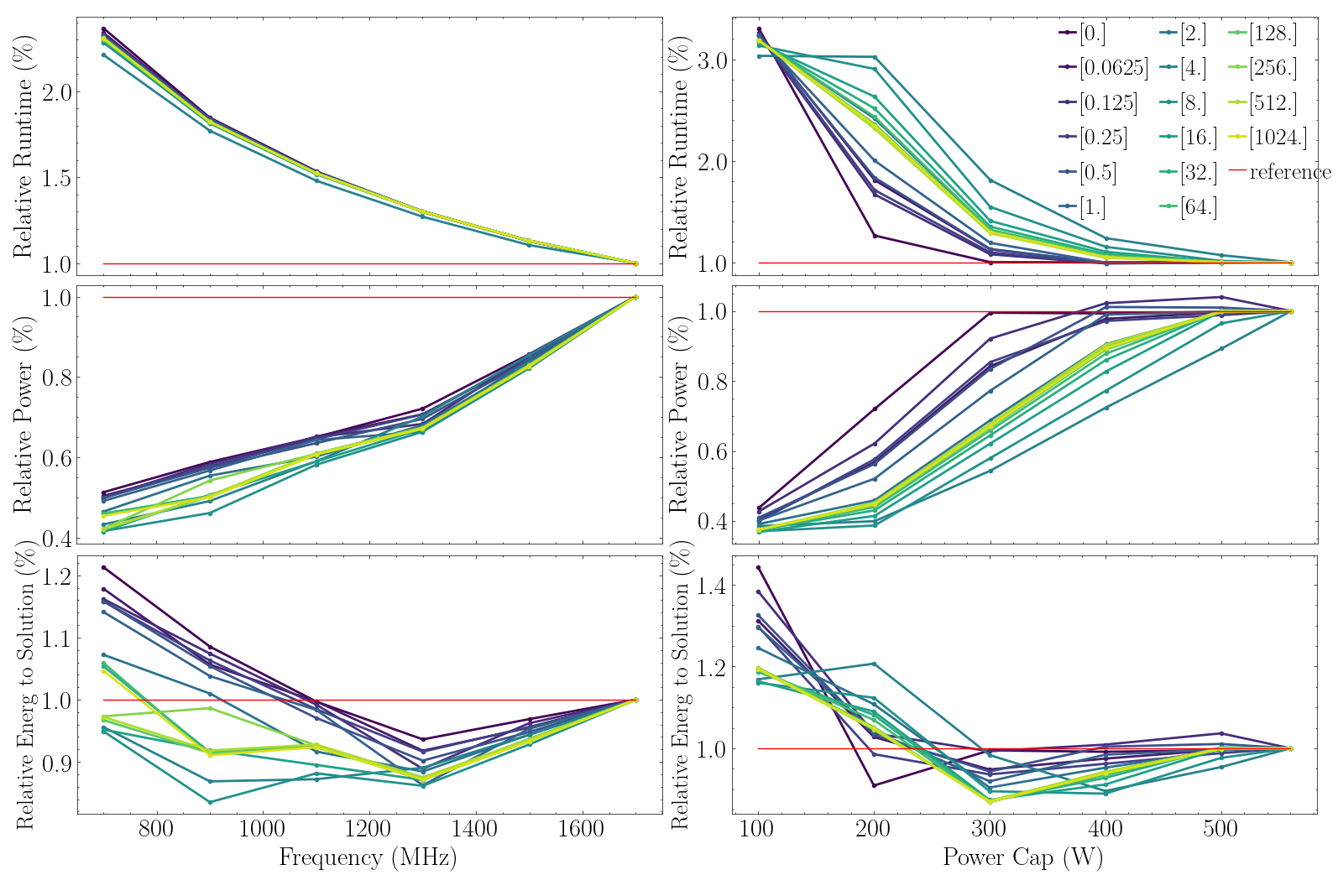}
    \caption{VAI Plot: Left fixed frequency (700MHz -- 1700MHz, Right Power Cap (100W-560W).
    For each we show: Runtime (top), power used (mid), and energy to solution (bottom).
    The values are normalized to 1\@.0, representing the uncapped case at 1700MHz/560W respectively. Each line is a specific arithmetic intensity from 1/16 to 1024 in powers of two. The arithmetic intensity of 0 is a stream copy call.}
    \label{fig:roofline_Freq_PowerTimeEnergy}
\end{figure}

In addition to the measured performance and power, Figure~\ref{fig:roofline} also recorded the slowdown in terms of time to completion with different power management settings. Since the benchmark is designed to provide a fixed number of operations of a specific operational intensity, we present the time to solution relative to the normal run (maximum frequency or no power cap) and derive the energy to solution in Figure~\ref{fig:roofline_Freq_PowerTimeEnergy}. %displays, from top to bottom, power, time, and energy to solution. The left side shows experiments based on a fixed frequency setting, while the right side shows experiments based on a power cap. 
In both frequency and power settings, we see that limiting either the frequency or the power provides us with an improvement in energy-to-solution under the trade-off of runtime. In the case of frequency limit, the most consistent energy-to-solution is observed at 1300\,MHz, with the time trade-off of averaged to 30\%. For the power-limited case, power limits lower than 300 watts see a significant increase in runtime, and the higher power caps do not impact the application enough to save power. As we see from Figure~\ref{fig:roofline},  the power consumption of the code peaks at an arithmetic intensity of $4$, reaching the \gls{TDP} of 560. At any other arithmetic intensity, even the higher ones, peak power is not consumed. It is important to note that a power limit only affects codes surpassing the limit, while a set frequency affects all. %As a reminder a power limit only affects codes surpassing the limit, while a set frequency affects all. 
With these insights, we summarized the power consumption, runtime increase, and energy saving by averaging across all arithmetic intensity for each frequency and power setting in Table~\ref{tab:gpu_power_saved}.

%\subsection{GPU Characteristics for Cache and Memory Latency}
\subsection{Power Characteristics of GPU Memory and Caches}
%\subsection{GPU Memory Bandwidth Benchmark}

The performance of the GPU also depends on from which layer the GPU will retrieve the data for computation. Low latency cache layers such as L1 and L2 provide high bandwidth, which keeps the utilization of ALU high. However, in real-world applications, it is often not possible to find the requested data in cache layers, and the GPU devices need to fetch the data from the memory layer. It could be due to various reasons, such as the big data size, a non-contiguous or strided memory access pattern, or an algorithm retrieving the data randomly.
\begin{figure}[t]
  \centering
  \includegraphics[width=\linewidth]{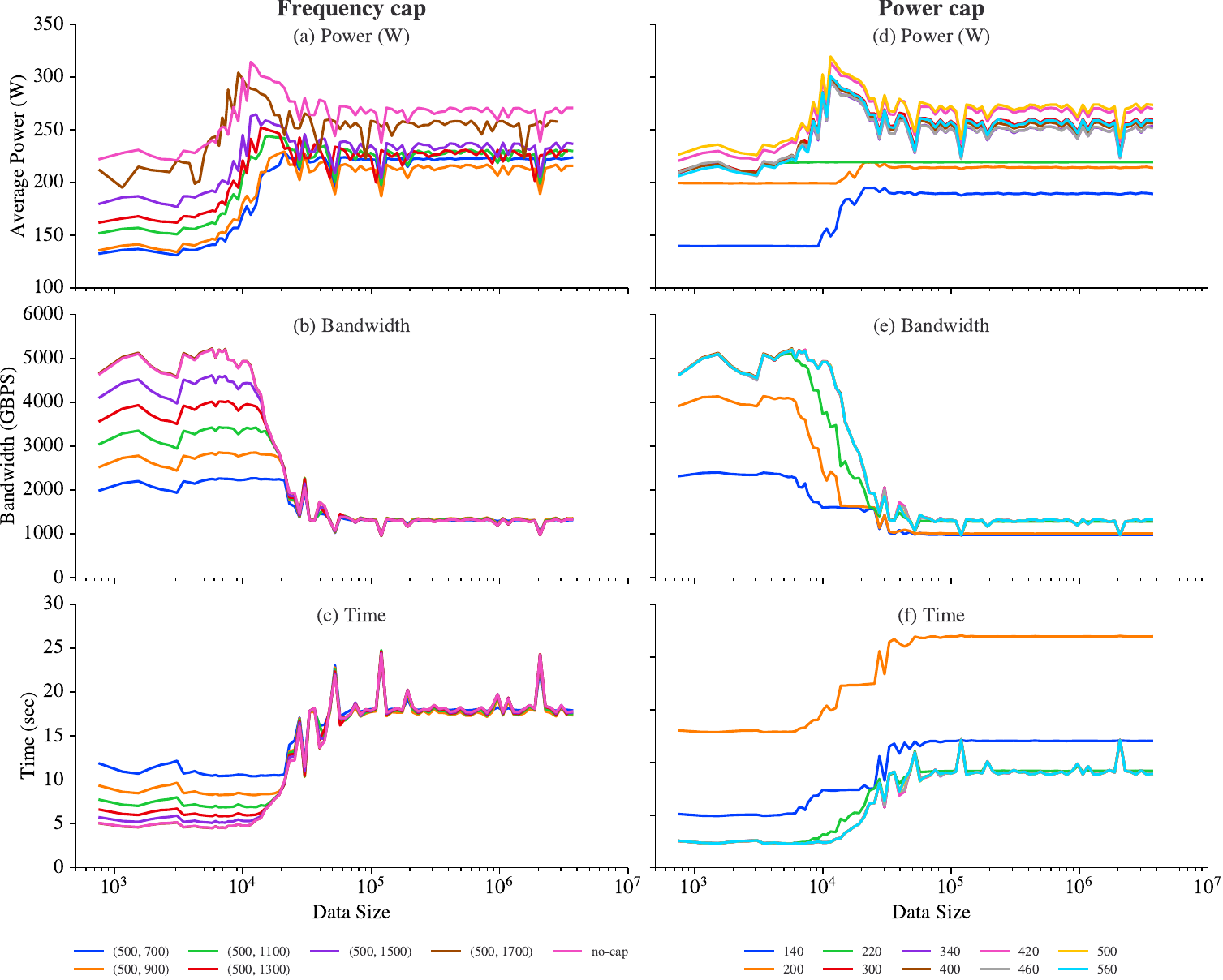}
    \caption{GPU characteristics for memory intensive benchmarks. The x-axis represents the size of the input data. Each curve represents a different frequency band (left) and operating power (right). }
  \label{plots:gpu_benches_analysis}
\end{figure}

%% Template for a single column-width table
\begin{table}
  \centering
  \caption{Percentage of the average power and runtime for VAI and memory bandwidth (MB) benchmark for \texttt{(a)} varying frequency cap and \texttt{(b)} for varying power cap. Values of VAI is average across the arithmetic intensity.
  }
  \label{tab:gpu_power_saved}
  %% Footnotesize if a much smaller font is needed
  \footnotesize
  \setlength\tabcolsep{3pt}
%%  \begin{tabular}{ccccc}
%%    \toprule
%%      Mode & \multicolumn{2}{c}{$\Delta$ Avg. power(\%)} & \multicolumn{2}{c}{$\Delta$ Avg. runtime increase(\%)}\\ 
%%    \midrule
%%        & VAI & MB & VAI & MB\\
%%        \cmidrule{2-5}
%%      1700 &  & 12  &  & 0 \\
%%      1500 &  & 12  &  & 0 \\
%%      1300 &  & 15.2 &  & 0.01  \\
%%      1100 &  & 15.1 &  & 0.01  \\
%%       900 &  & 20  &  & 0.02  \\
%%       700 &  & 19  &  & 0.01  \\
%%\bottomrule
%%\end{tabular}
  \begin{tabular}{ccccccc}
    \multicolumn{7}{c}{(a) Frequency Cap}\\
    \toprule
      \multirow{2}{*}{\rotatebox[origin=c]{0}{\parbox[c]{1cm}{\centering }Freq cap}} & \multicolumn{2}{c}{Avg. power(\%)} & \multicolumn{2}{c}{runtime increase(\%)} & \multicolumn{2}{c}{Avg. energy used(\%)}\\ 
    \cmidrule{2-7}
        & VAI & MB & VAI & MB & VAI & MB \\
        \midrule
      1700 & 100  & 100   & 100  & 100 & 100 & 100\\
      1500 & 83.7 & 87.2   & 112.8 & 99.7 & 94.4 & 86.9\\
      1300 & 68.2 & 84.5 & 129.8 & 99.5 & 88.6 & 84.3 \\
      1100 & 61.8 & 84.9 & 152.2 & 98.9 & 94.0 & 83.8\\
       900 & 53.3 & 79.7   & 182.4 & 99.0 & 97.3 & 79.7 \\
       700 & 46.0 &  82.9   & 231.0 & 99.1 & 106.3 & 95.7  \\
  \bottomrule
\end{tabular}
\begin{tabular}{ccccccc}
    %\\
    \multicolumn{7}{c}{(b) Power Cap}\\
    \toprule
      \multirow{2}{*}{\rotatebox[origin=c]{0}{\parbox[c]{1cm}{\centering }Power cap}} & \multicolumn{2}{c}{Avg. power(\%)} & \multicolumn{2}{c}{runtime increase(\%)} & \multicolumn{2}{c}{Avg. energy used(\%)}\\ 
    \cmidrule{2-7}
        & VAI & MB & VAI & MB & VAI & MB \\
        \midrule
      560 & 100  & 100 & 100   & 100& 100 & 100\\
      500 & 99.3 & 100 & 100.4 & 99.9& 99.7 & 92.2\\
      400 & 90.8 & 99 & 105.2 & 100.1& 95.0 & 93.6 \\
      300 & 72.7 & 99 & 128.4 & 100& 91.3  & 94.7\\
      200 & 49.3 & 85 & 222.3 & 125.7& 105.7 & 84.6\\
      %100 & 39.0 & 74 & 319.4 & 133.6 & 124.7 & 90.1\\
  \bottomrule
\end{tabular}
\end{table}

The plots in Figure~\ref{plots:gpu_benches_analysis} show the performance of GPU by varying management knobs. 
%In this benchmark, we perform strided access for varying data sizes. 
We primarily focus on frequency and power capping aspects of the GPU in fetching data from L2-cache layer and High-Bandwidth (HBM) GPU memory
layer. The left column of Figure~\ref{plots:gpu_benches_analysis} illustrates average power, bandwidth, and time to completion plots for different frequencies. Similarly, the plots in the right column are under different max operating power for varying input data sizes.

For frequency cap runs, we see that when the size of the data is less than 16 MB (size of L2-cache), the lower the frequency cap, the lower the bandwidth, and the higher the runtime for a benchmark. However, when the data size increases beyond 16 MB size, we see that increasing the frequency cap has no effect on the performance and thus run time remains almost similar for all runs. This shows that the workloads that have frequent memory access are not impacted by lowering GPU frequency because they are bounded by HBM memory bandwidth. On the other hand, applications that have the data in the cache or always have cache hits, are dependent on GPU frequency and their performance drops, run time increases as the max frequency cap drops.

For the power cap runs, we observe a similar pattern when the data fits in the L2-cache the drop in maximum power results in a drop in performance and longer runtime. The smaller value of the power cap, i.e. 140 W and 200 W have lower performance even when the data size is more than the L2-cache and resides in memory. The lower power cap reduces the performance even when the workload is memory-bound. When the data fits in the cache, the power usage is strictly below the max power cap shown by horizontal curves. However, when the data is accessed from HBM memory, it costs additional power, and the max power cap limit is breached shown by power curves associated with max power cap of 140 W and 200 W, illustrated by orange and blue curves in Figure~\ref{plots:gpu_benches_analysis}(d). We summarize the power consumption, runtime increase, and energy saving for each frequency and power setting alongside the previous VAI benchmark data in Table~\ref{tab:gpu_power_saved}. 
The values of maximum frequency and power cap are at 100\% for average power, runtime, and average energy. For lower frequency and power cap,  the relative values to the maximum capping are shown. 
We used the energy savings percentage from Table~\ref{tab:gpu_power_saved} for estimating energy savings in Section~\ref{sec:Projection}(c).
%\subsection{Case Study: Memory Bound Application}
%\subsection{: Frontier Application Performance}
\subsection{A Case Study: Power Characteristics of GPU-based Louvain Community Detection Application on Frontier}
\label{sec:Analysis_Graph_App}
%For an application, 
\begin{figure}[t]
  \centering
  \includegraphics[width=\linewidth]{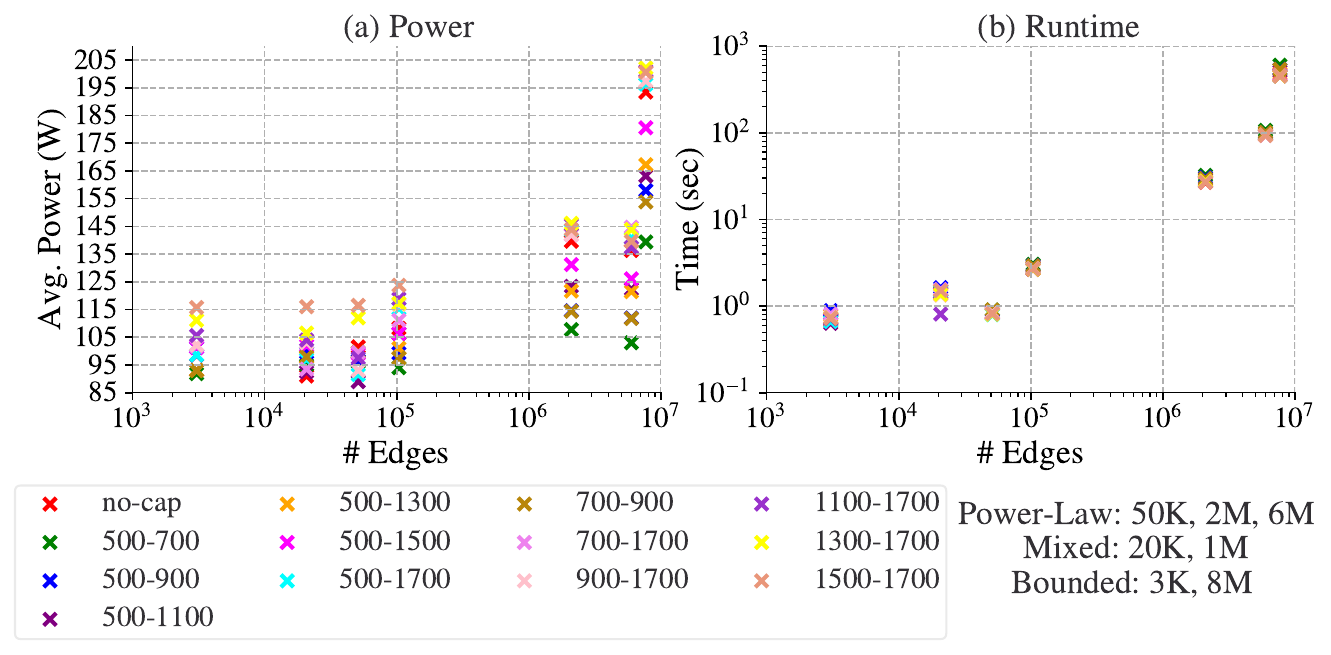}
    \caption{Characteristics of graph based real HPC workload at different frequencies}
  \label{plots:graph_perf_freq}
\end{figure}
Moving on from the benchmark results, we are exploring the real application region to verify if our characterization of energy-saving changes applies to real-world applications. We selected a graph application to expand the idea of slowing down GPU ALU speed while in a memory-bounded region. 

We analyze the performance of the graph application described in Section~\ref{subsec:benchmarks}-c in varied frequencies and power caps. %We observe a similar trend for both frequency and power and thus have 
We have shown the performance for frequencies in Figure~\ref{plots:graph_perf_freq}. We observe that, in the case of the social networks (representatives of power-law graphs), the runtimes are less sensitive to frequencies compared to a road network ($d_{max}=9$, $d_{avg}=2$) that is a bounded-degree graph. In the lower frequency ranges ($500$MHz, $700$MHz), the performance is impacted more compared to the higher frequency ranges. The bounded-degree graphs are more sparse and expected to have less power usage. For this application, the workload is distributed in such a way that for higher degree-based networks, the computation of a vertex's community assignment is handled by a group of threads within a wavefront or the full wavefront. For the sparse networks, a single thread completes the operation. The workload being much imbalanced for this example network, we see more power usage and the performance is also sensitive to the frequency. For the social networks, we observe a more balanced workload, and the performance is less sensitive to frequency changes. For the largest three networks (8M, 6M, 2M), we observe an energy saving of ($5.23\%$, $2.91\%$,$3.32\%$) with at most $5\%$ increase of runtime at $900$MHz. 
%For power caps, we see similar behavior for the power-law and bounded networks.
\par
In Figure~\ref{plots:graph_perf_freq}(a), we see the maximum power value for the 8M road network is up to $205$W. So, this input network when power-capped at $140$W or $180$W, shows a runtime increase of $35.8\%$ and $7.1\%$ respectively. However, the energy savings is $11\%$ and $26\%$ respectively. We can see up to $15\%$ energy saving when power is capped up to $220W$ with no increase in runtime. Since the graph application falls into both memory and compute-intensive regions, supports the energy savings findings shown for both VAI and MB benchmarks given in Table~\ref{tab:gpu_power_saved}.

%The GPU micro-benchmark is directed toward understanding the GPU effective bandwidth for non-contiguous memory access patterns under varying frequency ranges and power capping. 

%The plots in Figures~\ref{freq cap} and~\ref{power cap} show the effect of frequency and power cap on the performance of the benchmark applications. 
%For this task, we refer to the earlier works on micro-benchmarks~\cite{gpu_benches:ernst2021performance, gpu_benches:ERNST2023152}. 

% \begin{figure}[t]
%   \centering
%   \includegraphics[width=\linewidth]{plots/template_single.pdf}
%     \caption{Roofline performance plot for Summit and Frontier GPUs}
%   \label{plots:example}
% \end{figure}

%\section{TBD:w Power Decomposition + Modalities + GPU Characterization }
\section{From Benchmark to Power Projection at System Scale }
\label{sec:Projection}
%\ahmad{list: characterization, decomposition and projection}
We now perform the mapping step of modal analysis and mapping as outlined in the method Section~\ref{subsec:method_modal_decomposition} and Section~\ref{subsec:method_mapping_the_modes}.
The results from the GPU benchmarks enable us to estimate the savings projection of frequency and power capping for real-world HPC applications. We first characterize GPU power utilization collected over three months on the Frontier system. We leverage the benchmark results of VAI or compute-intensive (CI) and memory-intensive (MI) benchmarks to decompose the real-world power distribution into four bins corresponding to resource utilization. The results allowed us to develop an energy efficiency method and estimate its impact on the Frontier system.  
\subsection{Characterization of Frontier Workloads}
\label{sec:Modalities}
\begin{figure}[t]
  \centering
  \includegraphics[width=0.85\linewidth]{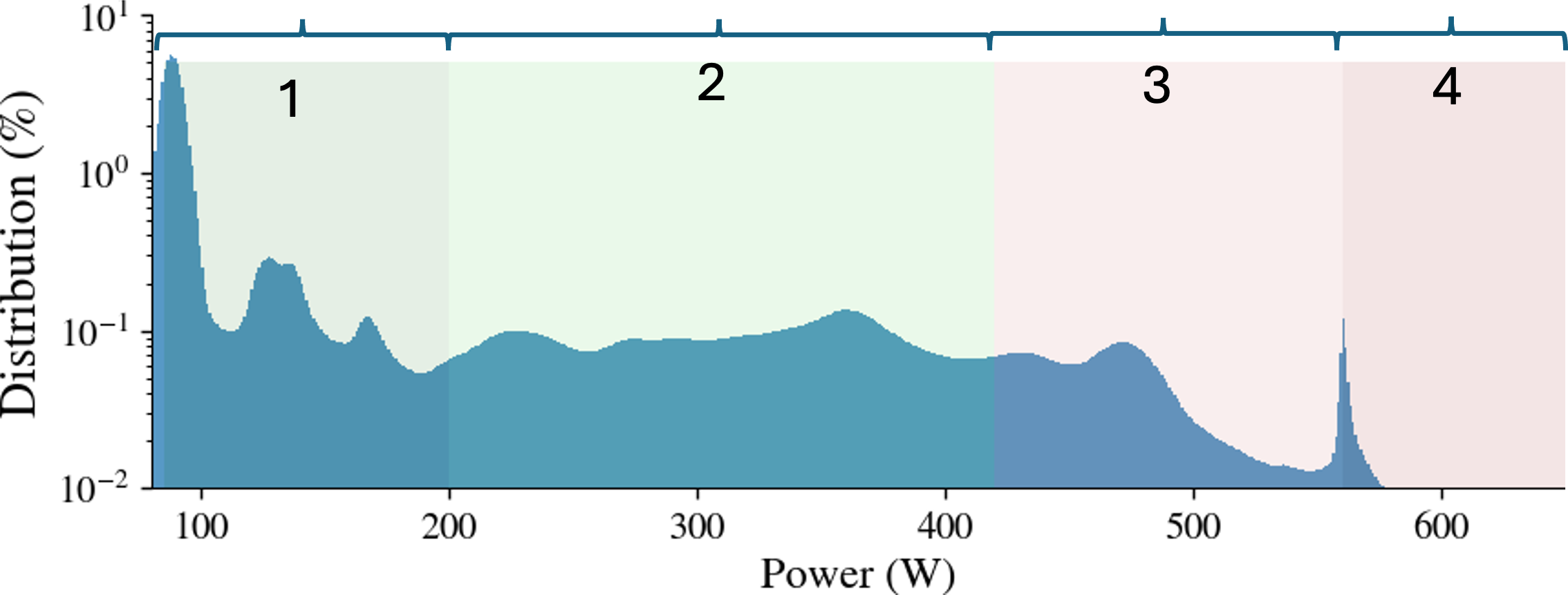}
    \caption{Frontier system-wide distribution of GPU power utilization.}
  \label{plots:frontier_gpu_distribution}
\end{figure}

We characterize the HPC workload power utilization to gain insight and develop an understanding of GPU behavior for jobs exhibiting diverse resource utilization patterns. The plot in Figure~\ref{plots:frontier_gpu_distribution} shows the distribution of GPU power utilization of all the jobs executed on Frontier over a timeframe of 3 months.
Peaks or local maxima in the plots show the region or zone of operation frequently exhibited by jobs, which corresponds to the actual GPU resource utilization by jobs. The distribution of the data points shows that several peaks are close to low power utilization and few peaks towards higher power utilization. 
The idle power of a GPU is between $88$ to $90$ W. The low power utilization shows that the GPUs are constrained by non-compute resources and the high-power utilization shows that the GPU loads are CI or flop intensive. 

\begin{figure}[t]
  \centering
  \includegraphics[width=0.85\linewidth]{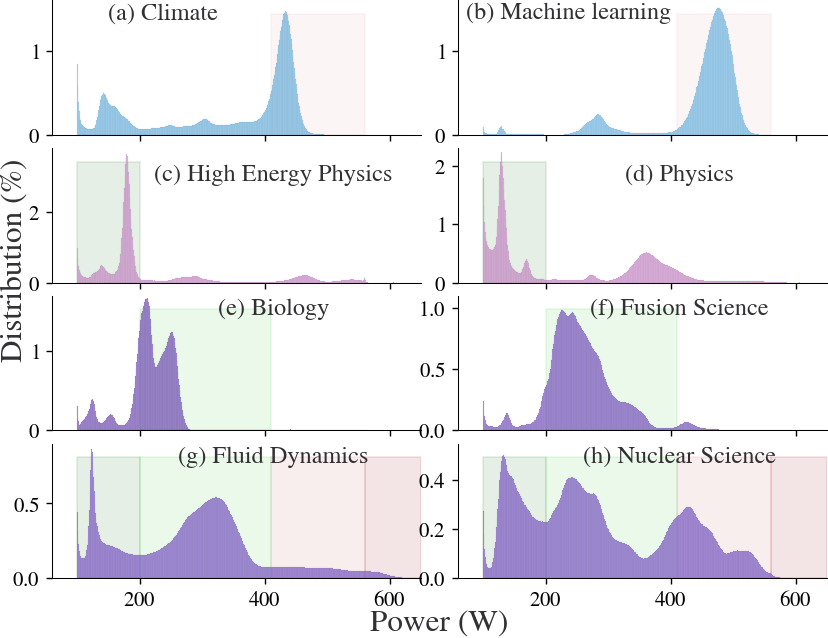}
    \caption{Characterization of workloads based on GPU uitilization for science domains.Shaded region shows the range of operation as described in Table~\ref{tab:gpu_modes}}
  \label{plots:domain-science-distribution}
\end{figure}

%The distribution of power in Figure~\ref{plots:frontier_gpu_distribution} although shows peaks but modes are not clearly seen because the data-set contains all jobs and their distributions are diffused into one another. 
To better understand the resource utilization and the GPU power profile, we plot the distribution of GPU power utilization science domain-wise. We derive science-domain from \texttt{project\_id} field in the job scheduler log, where the prefix of each \texttt{project\_id} is equivalent to science-domain. 
The power distributions in Figure~\ref{plots:domain-science-distribution}, shows the GPU power utilization of some of the domains, and the modalities are clearly seen within each-science domain. 
This is because the science domains most often run applications that have similar workloads that uses specific GPU resources. We see that in Figure~\ref{plots:domain-science-distribution} (a) and (b), jobs are primarily high power utilization and correspond to applications that are compute-intensive and are utilizing all the GPU hardware resources close to optimal. 
The plots in Figure~\ref{plots:domain-science-distribution} (c) - (d) exhibit low power values, which points to under utilization of GPUs' compute resources. 
These applications are latency bound, i.e. non-compute resource bound applications leading to fewer compute and memory resource utilization and thus lower GPU power utilization. 
The plots in Figure~\ref{plots:domain-science-distribution} (e) and (f) represent applications that show power utilization in the range between the two extreme power ranges. 
These are memory-bounded applications that exhibit sub-optimal compute resource utilization but still significant enough usage. 
Then there are other science domains represented by plots in Figure~\ref{plots:domain-science-distribution} (g) and (h) in which the applications run in multiple zones of operation across job runtime. 
The disaggregation of system-wide GPU distribution into the science domain shows that GPU power usage represent GPU resource utilization and the nature of workloads.

\subsection{Decomposition of Power Distribution in Modes}
%\subsection{Understanding power-profile}
%In this section we leverage the GPU benchmark analysis for understanding HPC application resource utilization. 
%As shown 
%We group the GPU utilization in four different ranges or zones of operation based on power values. 
%Understanding high-level trends in GPU power pattern
%In this section we observe the distribution of GPU power across the system-wide workloads. We want to observe patterns exhibited by different jobs and science domains.  
%\input{plots/understanding_gpu_operation}
Utilizing the distribution of real-world HPC workloads and the GPU power patterns exhibited by different resource utilization benchmarks, we group the different regions of workloads based on power values from benchmark results. 
We highlight the regions in Figure~\ref{plots:frontier_gpu_distribution} and quantitatively describe them in Table~\ref{tab:gpu_modes}.
The region of compute- and memory-intensive operations is the zone of interest for our work. In these two regions, our benchmarks showed saving opportunities. Boosted frequency is the power range above 560 W. A GPU can only sustain in this region for a short duration, before falling back to a frequency level that corresponds to less than or equal to 560 W of power. In the benchmark section, we measured steady-state power consumption, which is longer than boost duration, while with the sampling of the telemetry collection over the months, we observed values in the boost frequencies. The boosted frequency region covers only $1\%$ of the GPU hours but we do not record it in benchmark and thus cannot do the power savings projection for it. 
For latency, network \& I/O bound regions we did not observe any savings opportunities. The frequency or power cap on the benchmark during this phase of operation proportionally raised the runtime with a decrease in power. Thus, no benefits in the energy-to-solution, but the time-to-solution was significantly increased.
As illustrated above in Section \ref{sec:gpu_characterization_analysis}-A, during the benchmark analysis of compute-intensive workloads GPU derives the highest amount of power which is between 420-560 W. Likewise, in Section \ref{sec:gpu_characterization_analysis}-B, the workloads during the memory-intensive operations derive power values around 200-420W, thus we have marked that region as memory intensive operation regions. 
%Low power utilization is marked as either latency bound or network bound applications. Latency or network bound regions are marked between 90-200W where 90 corresponds to approximate idle power value. We call the region beyond 520 W as boosted frequency region of operation activated by GPU momentarily.
%% Template for a single column-width table
\begin{table}
  \centering
  \caption{Leveraging GPU modalities for Resource Utilization}
  \label{tab:gpu_modes}
  %% Footnotesize if a much smaller font is needed
  \footnotesize
  \setlength\tabcolsep{4pt}

  \begin{tabular}{clcc}
    \toprule
     Region& Mode (region of operation) & Range (W) & GPU Hrs. (\%)\\ 
    \midrule
    1& Latency, Network \& I/O bound & $\leq 200$ & 29.8\\     
     2& Memory intensive (M.I.) & 200-420 & 49.5\\ 
     3& Compute intensive (C.I.) & 420-560 & 19.5 \\
    4& Boosted frequency & $\ge 560 $ & 1.1\\     
  \bottomrule
\end{tabular}
\end{table}

Although boundary regions may be diffused into one another and may not be well defined, based on the benchmark results the order of the zone classification is accurate. This analysis enables us to have an insight into the savings opportunities and provides us with an estimate that is closer to the actual savings. The classification of data into zones gives a high-level view of the behavior of the workloads, resource utilization, and a quantitative value for projecting saving estimates. %Furthermore, the science domain  

\subsection{Performing the Energy Savings Projection at System Scale}
%Software Driven Energy Efficiency and Impact on HPC}
\label{sec:SavingsProjection}

%% Template for a single column-width table
\begin{table}
  \centering
  \caption{Estimated energy savings when frequency and power capped applied at system-wide.}
  \label{tab:energy_saving_summary}
  %% Footnotesize if a much smaller font is needed
  \footnotesize
  \setlength\tabcolsep{3pt}
  \begin{tabular}{cccccccc}%||lcc}
  \multicolumn{8}{c}{(a) Frequency Cap}\\
    \toprule
      Total &  Freq. & C.I.  & M.I. & T.S. & Energy & $\Delta T$ & Energy \\
      Energy& (MHz) & (MWh) & (MWh) & (MWh) & Savings & Time &Sav. (\%)\\
     Used&&&&&(\%)&(\%) & $\Delta T=0$\\
    \midrule      
       \multirow{ 4}{*}{\rotatebox[origin=c]{90}{\parbox[c]{1.5cm}{\centering 16820~MWh}}} & 1500 & 115.3 & 928.2& 1043.5 & 6.2 & 1.7&5.5\\
        & 1300 &234.7 & 1112.4& 1347.1& 8.0 & 4.1&6.6\\
        & 1100 &123.5& 1154.9& 1278.4 & 7.6 &7.1 &6.8\\
        & 900 &55.6 & 1438.3& 1493.9 & 8.8 &11.2 &8.5\\
        & 700 & -129.7 & 304.6 & 174.9 & 1.0 &17.7 & 1.8\\
    \midrule
\end{tabular}
  \begin{tabular}{cccccccc}
  \\
  \multicolumn{8}{c}{(b) Power Cap}\\
    \toprule
      Total&  Power & C.I.  & M.I. & T.S. & Energy & $\Delta T$& Energy\\
     Energy & (W) & (MWh) & (MWh) & (MWh) & Savings & Time &Sav. (\%)\\
     Used&&&&&(\%)&(\%)&$\Delta T=0$\\
    \midrule      
       \multirow{4}{*}{\rotatebox[origin=c]{90}{\parbox[c]{0.3cm}{\centering 16820 MWh}}} & 500 & 6.17 & 552.65& 558.83 & 3.32  & 0.1&3.2\\
        & 400 &102.96& 453.46& 556.42 & 3.30  & 0.7&2.6\\
        & 300 &179.16 & 375.52& 554.68 & 3.2  & 3.83&2.2\\
        & 200 &-117.38 & 1091.14& 973.75 & 5.7 & 16.53&6.4\\
    %\bottomrule
    \midrule
\end{tabular}
\end{table}

%% Template for a single column-width table
\begin{table}
  \centering
  \caption{Estimated energy savings by frequency capping when applied on 6 domains that has at-least one cell in red color and the large jobs of size  A, B, \& C.}
  \label{tab:energy_saving_summary_selected_domains}
  %% Footnotesize if a much smaller font is needed
  \footnotesize
  \setlength\tabcolsep{4pt}
  \begin{tabular}{cccccccc}%||lcc}
  %\multicolumn{8}{c}{(a) Frequency Cap}\\
  %\\
    \toprule
      Total &  Freq. & C.I.  & M.I. & T.S. & Energy & $\Delta T$ & Energy \\
      Energy& (MHz) & (MWh) & (MWh) & (MWh) & Savings & Time &Sav. (\%)\\
     Used&&&&&(\%)&(\%) & $\Delta T=0$ \\
    \midrule      
          \multirow{ 4}{*}{\rotatebox[origin=c]{90}{\parbox[c]{0.5cm}{\centering 16820 MWh}}} & 1500 & 92.79 & 716.75 & 809.55  & 4.8 & 1.8 & 4.2\\
        & 1300 & 188.90 & 859.01 & 1047.91 & 6.2 & 4.2 & 5.1\\
        & 1100 & 99.42 & 891.84 &  991.26 & 5.8 & 7.3 & 5.3\\
        & 900 &  44.74 & 1110.70 & 1155.44 & 6.8 & 11.5 & 6.6\\
    \midrule
\end{tabular}
% \begin{tabular}{cccccc}%||lcc}
%   \\
%   \multicolumn{6}{c}{(b) Power Cap}\\
%     \toprule
%      Energy &  Power & C.I.  & M.I. & T.S. & Energy \\
%      Used & (W) & (MWh) & (MWh) & (MWh) & Savings (\%)\\
%      %& C.I. & M.I. \\     
%     \midrule      
%        \multirow{ 4}{*}{\parbox[c]{0.5cm}{\centering 16820  MWh}} & 500 & 4.97 &426.77 & 431.74 & 2.5  \\
%         & 400 &82.85&350.17 & 433.02 & 2.5 \\
%         & 300 &144.16 & 289.98& 434.14 & 2.5 \\
%         & 200 &-94.45 & 842.60& 748.14 & 4.4 \\
%     %\bottomrule
%     \midrule
% \end{tabular}
\end{table}

\begin{table}
  \centering
  \caption{Job scheduling policy of Frontier system}
  \label{tab:sched_policy}
  %% Footnotesize if a much smaller font is needed
  \footnotesize 
  \begin{tabular}{ccc}
    \toprule
      Job size & Num-nodes & Max. Walltime (Hrs.)\\ 
    \midrule
      A  & 5645 - 9408 & 12\\
       B & 1882 - 5644 & 12\\
       C & 184 - 1881 & 12\\
       D & 92 - 183 & 6 \\
       E &  1 - 91 & 2\\
  \bottomrule
\end{tabular}
\end{table}

We estimate the energy savings on
Frontier workloads based on the findings from
the GPU benchmarks and analysis of GPU utilization. We estimate the potential savings when
frequency or power capping is applied to GPUs. We used
the data of the percentage of average power used with respect
to power used for maximum frequency and maximum power, as shown in
Table~\ref{tab:gpu_power_saved}. 
To calculate an energy-saving estimate at a system-wide scale, out of the four regions of operation, we only select the data points from compute- and memory-intensive zones (Table~\ref{tab:gpu_modes}) that showed energy savings during the benchmark runs.

The results of the frequency-cap and power-cap for system-wide energy savings are shown in Table~\ref{tab:energy_saving_summary}(a) and ~\ref{tab:energy_saving_summary}(b), respectively. For each frequency cap and power cap, we show the savings for compute-intensive (CI) or region 2, for memory-intensive (MI) or region 3, combined total savings (TS), energy savings in percentage calculated as $\texttt{TS}/\texttt{total energy used}(1680)$, percentage increase in run time ($\Delta T$), and energy savings when runtime is not increased ($\Delta T=0$). The GPU energy savings are shown over the complete dataset for the 3-month period and the total GPU energy used in that time was $16820 MWh$. 
The energy savings estimates in Table~\ref{tab:energy_saving_summary} are calculated when the power management caps are applied across the system and independent of science domains. It shows applying a frequency cap to applications provides maximum potential savings.
%However, it will be more efficient and practical to apply a frequency cap to applications that gives maximum savings benefits. 

\begin{figure}[t]
  \centering
  \includegraphics[width=\linewidth]{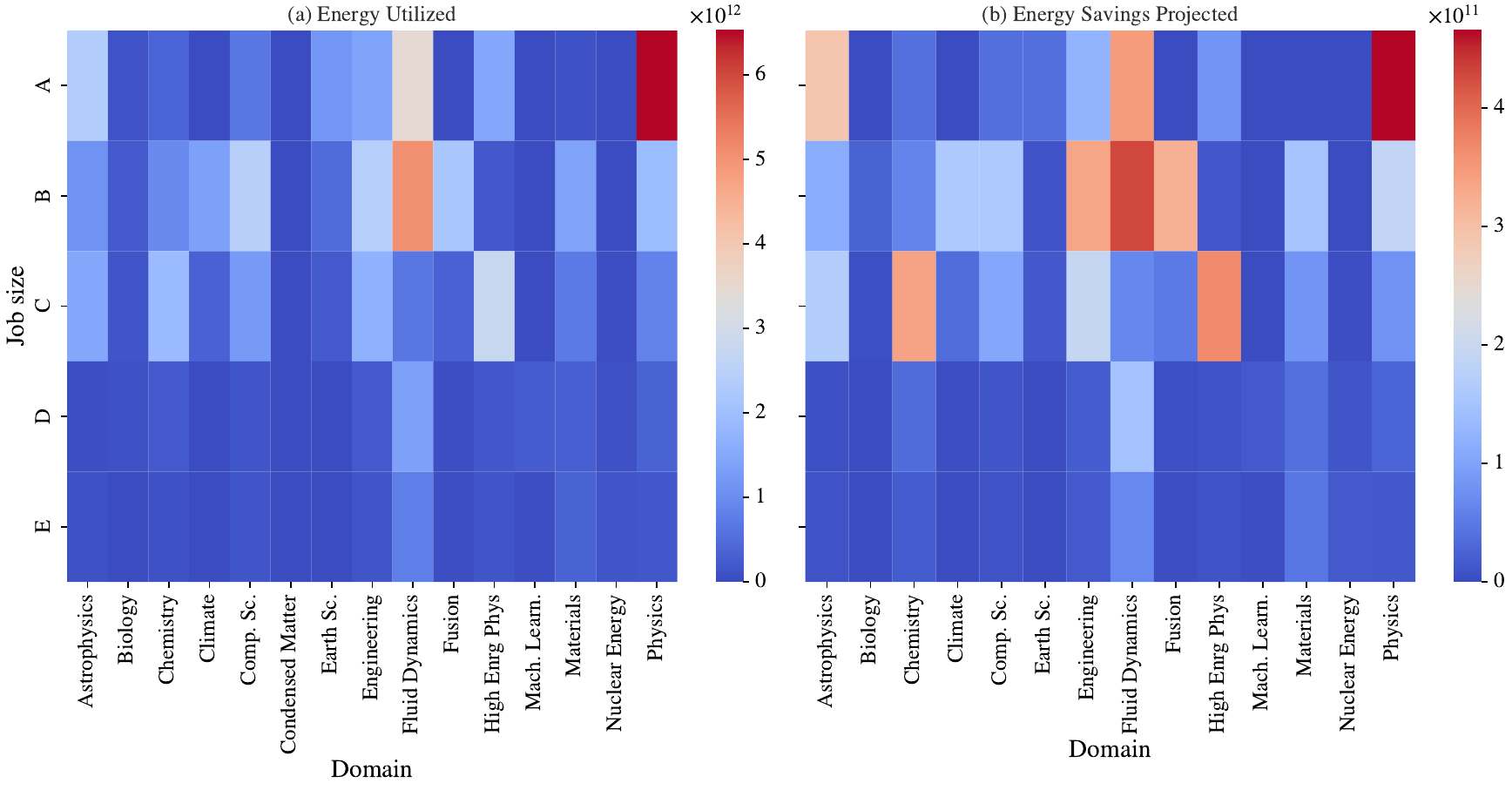}
    \caption{Figure shows the heatmap of total energy utilized and energy saved by science domains vs size of the jobs.}
  \label{plots:energy-saved}
\end{figure}

The heatmaps in Figure~\ref{plots:energy-saved} show the domain-wise and job-size-wise total GPU energy usage in~\ref{plots:energy-saved}(a) and estimated energy savings in ~\ref{plots:energy-saved}(b) if a frequency cap of 1100 MHz is applied. Job size is defined based on the Frontier scheduling policy listed in Table~\ref{tab:sched_policy}. Frontier is a leadership-scale HPC system geared towards large workloads, which is why large jobs such as jobs A and B run on a large number of nodes and have more wall times, as shown in Table~\ref{tab:sched_policy}, and thus have longer GPU hours. It can be observed that most of the science domain primary energy utilization comes from jobs that belong to job sizes A and B. 
%It can also be seen from 
Figure~\ref{plots:energy-saved}(b) also illustrates that most of the energy savings estimates are from jobs in A, B, and C, in other words, from significantly large jobs.

From the domain-wise view (Figure~\ref{plots:energy-saved}(b)), we observe that a few domains and large jobs show significantly more energy savings. We present the estimated energy savings only for a few selective science domains that exhibit at least one cell in the red color shade in Figure~\ref{plots:energy-saved}(b), and from job-size group \texttt{A}, \texttt{B}, and \texttt{C}.
Frequency capping saving results are shown in  Table~\ref{tab:energy_saving_summary_selected_domains} and it can be observed that it is a significant percentage of the results in Table~\ref{tab:energy_saving_summary}. Based on our results, some applications have more savings benefits than others.

%%%%%%%%%%%%%%%%%%%%%
%% Discussion
\section{Discussion}
\label{sec:discussion}

With the results from Section~\ref{sec:methodology}, Section~\ref{sec:gpu_characterization_analysis} and Section~\ref{sec:Projection} we can quantify an upper ceiling for potential savings.
This is essential to justify or dismiss development efforts for centers with power constraints and energy priorities.

We project potential energy savings of up to about $8.5\%$ without a performance slowdown, while the savings increase to up to about $8.8\%$ if a performance penalty of $11.2\%$ is tolerated (Table~\ref{tab:energy_saving_summary}). We also demonstrated that the power management methods need not be applied at the system scale but can be applied to selected domains and job sizes to still achieve significant savings
% benefits without affecting all HPC users
(Table~\ref{tab:energy_saving_summary_selected_domains}).

%To apply the presented method the power behavior of the system's nodes have to be characterized, while after the mo
%To apply the method developed, we first need to have a good understanding of performance impact and power characteristics, using benchmarks. The second step is projecting the results on to telemetry data.
%The telemetry data used and analyzed is currently limited to power. 
%Regarding benchmark insights, the derived knowledge from the GPU power characterization can be applied in a more nuanced manner.
%Aside from improving the insights gained from the benchmarks, 
The telemetry data can be augmented to include more precise application fingerprinting, with more precise sensitivity prediction regarding power management. However, HPC centers need to have the infrastructure to support huge data storage needs.
To provide a first up-ceiling for energy savings, our approach for projection can be used as a baseline to be lowered by an improved understanding of systems and opportunities. 

Regarding the limited view of telemetry data, a second aspect is that only the GPU power was evaluated. The other components are dwarfed ($<20\%$) by the GPU power consumption on a fully utilized node. A comprehensive power and resource utilization data would enable a more accurate estimate of energy saving. 
%To improve the prediction quality this has to be included. We conclude with the same note that this, again, would correct to potential savings opportunity down, making our approach a first upper ceiling.

%A major point in the discussion is, that this is a first approach of quantifying the energy savings on a system of leadership scale such as Frontier. 
%To improve the quality of prediction, the power management mechanisms have to be well understood,
%benchmarks that are reflecting usage characteristics and mapping these to the workload mix is essential.
%From the vendor side, providing not only the power controls such as DVFS and power limits, but a characterization of the different components of a GPU or compute component is essential. 
%%This was a major effort in this work, even if people claim that they know the ins and outs of a specific processor or accelerator already.

Additionally, based on technology developments, such assessments have to be re-evaluated to understand the tradeoffs and opportunities, which can be supported by the vendors.

%%%%%%%%%%%%%%%%%%%%%
%% Related work
%\input{txt/06_related}

%%%%%%%%%%%%%%%%%%%%%
%% Conclusion
\section{Conclusion}
\label{sec:conclusion}
We presented a method to estimate energy savings by applying power management techniques. We demonstrated the power savings during benchmark runs and a real HPC application run representing jobs utilizing diverse GPU resources. 
%Benchmark results showed the impact of frequency and power capping on GPUs by providing average energy savings of up to $15\%$. 
We used three months of HPC telemetry data to characterize power values based on resource utilization. Results from the benchmarks enabled us to group GPU utilization into four zones of operation and used power values as an estimation for GPU utilization. 
With this understanding of telemetry data and its classification in conjunction with power management, we calculated an upper bound of savings of up to $8\%$ of total GPU energy usage. 

%The vendor community can enhance the accuracy such projection efforts by not only providing power controls such as DVFS and power limits, but by providing processor characterizations.
%The vendor community From the vendor side, providing not only the power controls such as DVFS and power limits, but a characterization of the different components of a GPU or compute component is essential. 

This provides an important first step to the potential for quantifying the energy savings on a system of leadership scale such as Frontier.

%%% Reminder: Abstract:
%In this work we address the challenge of understanding the impact of software-driven power management mechanisms, as developed over the last decades. 
%We combine insights gained from benchmarking GPU to understand the power characteristics of applications with a telemetry data-driven approach to derive energy savings projection and apply it to the Frontier supercomputer at scale. Our findings highlight that, for three months of telemetry, resource-constrained jobs can save up to 8.5% of energy without performance penalty, which is equivalent to 1438 MWh. The contribution of this work is the method of deriving an upper limit for such best-case scenarios and applying it. The work provides energy-saving insights to understand and maximize the power/performance trade-off within constrained power budgets in the exascale era and beyond.

%\section*{Acknowledgment}
%We are grateful to the technical staff members at ORNL.

%% Instructions from the IEEE form
% The preferred spelling of the word ``acknowledgment'' in America is without 
% an ``e'' after the ``g''. Avoid the stilted expression ``one of us (R. B. 
% G.) thanks $\ldots$''. Instead, try ``R. B. G. thanks$\ldots$''. Put sponsor 
% acknowledgments in the unnumbered footnote on the first page.

\bibliographystyle{IEEEtran}
\bibliography{EnergyProjection}

\end{document}